\documentclass[12pt]{article}
\usepackage{amssymb,amsmath}

\newcommand{\dd}{\mbox{\rm d}}
\newcommand{\wg}{\wedge}
\newcommand{\gam}{\gamma}
\newcommand{\Gam}{\Gamma}
\newcommand{\dg}{\dagger}
\newcommand{\ddg}{\ddagger}
\newcommand{\tl}{\tilde}

\newcommand{\bgam}{\boldsymbol{\gamma}}

\newcommand{\cs}{$C$-space}

\newcommand{\DD}{\mbox{\rm D}}

\newcommand{\oo}{\over}
\newcommand{\p}{\partial}

\newcommand{\be}{\begin{equation}}
\newcommand{\bear}{\begin{eqnarray}}
\newcommand{\ear}{\end{eqnarray}}
\newcommand{\ee}{\end{equation}}
\newcommand{\lbl}{\label}
\newcommand{\bi}{\bibitem}
\newcommand{\ci}{\cite}

\newcommand{\vs}{\vspace}
\newcommand{\hs}{\hspace}

\begin{document}

\

\baselineskip .7cm 

\rightline{gr-qc/0507053}

\vs{27mm}

\begin{center}

{\LARGE \bf Spin Gauge Theory of Gravity in Clifford Space:
A Realization of Kaluza-Klein Theory in 4-Dimensional Spacetime}

\vs{3mm}

Matej Pav\v si\v c

Jo\v zef Stefan Institute, Jamova 39,
1000 Ljubljana, Slovenia

e-mail: matej.pavsic@ijs.si

\vs{6mm}

{\bf Abstract}

\end{center}

\vs{2mm}

A theory in which 4-dimensional spacetime is generalized to a
larger space, namely a 16-dimensional Clifford space ($C$-space)
is investigated. Curved Clifford space can provide a realization
of Kaluza-Klein. A covariant Dirac equation
in curved $C$-space is explored. The generalized Dirac field
is assumed to be a polyvector-valued object (a Clifford number)
which can be written as a superposition of four independent spinors,
each spanning a different left ideal of Clifford algebra. The
general transformations of a polyvector can act from the left
and/or from the right, and form a large gauge group which may
contain the group U(1)$\times$SU(2)$\times$SU(3) of the standard
model. The generalized spin connection in $C$-space has the
properties of Yang-Mills gauge fields. It contains the ordinary
spin connection related to gravity (with torsion), and extra parts describing
additional interactions, including those described by the
antisymmetric Kalb-Ramond fields.

\vs{8mm}

\section{Introduction}

Kaluza-Klein idea for the unification of the gravitational and electromagnetic
interaction by extending the dimensionality of spacetime is very
fascinating. In eighties it attracted much attention in its modern
formulation in which higher dimensional curved spacetimes $V_n$, $n>5$,
were considered. This enabled to treat Yang-Mills fields as
being incorporated in the metric tensor of $V_n$. When confronting the
theory with phenomenology, serious difficulties have arisen. In particular,
there is a notorious problem that a charged particle has an effective
4-dimensional mass which is of the order of the Planck mass. A search
for a realistic Kaluza-Klein theory has not been successful so far.

In the meantime the focus of attention has switched to string theory
(see e.g.\,\ci{Green})
which has turned out to be very promising in unifying gravity with 
gauge interactions. A lot of fascinating results have been obtained in the
last ten years, such as a discovery that string theory contains
higher dimensional objects (D-branes) and that there must be a single
underlying theory, the so called M-theory, which unifies different
known types of string theory (see, e.g., \ci{M-theory}, and references
therein). The presence of branes in string theory has
led to the revival \ci{BraneWorld} of the old idea \ci{BraneWorldOld} that
our 4-dimensional universe is a 4-dimensional surface, a world manifold
of a 3-brane, living in a higher dimensional spacetime 
(see also refs.\ci{PavsicBook,PavsicBrane,ReggeBrane}).

There is a number of works which further illuminate strings and branes from
the theoretical
point of view. For instance, Aurilia et al. \ci{Aurilia}, following the
original proposal by Schild \ci{Schild} and Eguchi\, \ci{Eguchi}, formulated
$p$-branes in terms of the tangent $(p+1)$-multivectors to a $p$-brane's
world manifold. So they obtained the quantum propagator of a bosonic 
$p$-brane in the quenched minisuperspace approximation \ci{AuriliaCastro}
which lead to a novel ---Clifford algebra based--- unified description
\ci{Castro,CastroFound,PavsicBook, AuriliaFuzzy,PavsicArena} of $p$-branes
for different
values of $p$. The background space that emerged in such a framework turned
out to be a Clifford space.

Since the seminal Hestenes's works \ci{Hestenes} Clifford algebra is
becoming increasingly popular in physics \ci{Lounesto}. It has been
realized that Clifford algebra is not only a useful tool for description
of the {\it known} geometry and physics, but it provides a lot of room
for {\it new}
geometry and {\it new} physics. As an intermediate step into that new direction
several authors have investigated, in a number of illuminating and
penetrating works \ci{CliffUnificOthers}--\ci{Crawford}, the idea
that Clifford algebra provides
a framework for unification of fundamental interactions. Common to all
those works is that they consider the Clifford bundle of spacetime, and so
the generators of Clifford algebra and
other related quantities depend on
position in {\it spacetime}. But other
researchers \ci{Pezzaglia, Castro, PavsicBook, PavsicSchladming,PavsicCliff,
PavsicArena,CastroPavsicReview,CastroPavsicHigher}
have formualted
a different theory which is based on the concept of {\it Clifford space}
(\cs). This is the space of oriented $r$-dimensional areas which, in the
case of flat $C$-space, can be described by {\it polyvectors} $X$,
superpositions of $r$-vectors, where $r$
takes the discrete values from 0 to 4. 
A position in $C$-space is described by sixteen coordinates
$x^M = (s,x^\mu,~x^{\mu \nu},~x^{\mu \nu \rho},
~x^{\mu \nu \rho \sigma})$ which are a generalization of the usual four
spacetime coordinates $x^\mu, ~\mu=0,1,2,3$. The role of Clifford space
as the arena for physics has been investigated in refs.
\ci{Pezzaglia, Castro, PavsicBook, PavsicSchladming,PavsicCliff,
PavsicArena,CastroPavsicReview,CastroPavsicHigher} .

If $C$-space is flat, we have essentially a generalization of special relativity
to the new degrees of freedom which reside in $C$-space. The new degrees
of freedom, i.e., the polyvector coordinates $x^M$, if associated with a
physical object, encode the information not only about the objects's center
of mass position, but also about its extensions and orientation.
A detailed ``shape" or configuration of the
extended object is not encoded by
$x^M$, only a partial information about the shape is encoded
\ci{PavsicArena}.
The extended objects are observed in 4-dimensional
spacetime, they exist in spacetime, and their extended nature is
``sampled" by the  coordinates $x^M$, which denote position in
16-dimensional $C$-space. The latter space is just like a multidimensional
configuration space associated with a many particle system which, of
course, still resides in spacetime. So we have both at once:
4-dimensional spacetime, and 16-dimensional \cs.

In particular, the extended objects can be just the fundamental (closed)
branes. It is well known \ci{Riesz,Teitler} (see also a recent systematic
exposition \ci{MankocIdeal} and refs.\,\ci{RodriguesIdeal} that the elements of
the right or
left minimal ideals of Clifford algebra can be used to represent spinors.
Therefore, a coordinate polyvector $X$ automatically contains not only
bosonic, but also spinor coordinates.
In refs.\,\ci{PavsicParis,PavsicSaasFee} it was
proposed to formulate string theory in terms of polyvectors, and thus
avoid using a higher dimensional spacetime. Spacetime can be 4-dimensional,
whilst the extra degrees of freedom (``extra dimensions") necessary for
consistency of string theory are in Clifford space.

In this paper we investigate the possibility that the arena itself is
to become a part of the play. We propose that Clifford space should be
considered as
 a dynamical, in general curved, space, analogous to the
spacetime of general relativity.
Since a dynamical (curved) Clifford space
has 16 dimensions,
it provides a realization of Kaluza-Klein
idea.  This approach has seeds
in refs.\, \ci{CastroPavsicConform, CastroPavsicHigher}, but explicitly
it was formulated in refs.\, \ci{PavsicParis, CastroUnified, PavsicUnified}.

We will first investigate some basic aspects of the classical general
relativity like theory in $C$-space. We point out that the geodetic equation
in $C$-space contains the terms that can be interpreted as being related
to the extra interactions (besides the ordinary gravitational one). 
Then we pass to quantum theory and discuss the concept of polyvector
valued wave function which can be written in a basis spanning four
independent left ideals of Clifford algebra.
We write the
Dirac-like equation in curved $C$-space by introducing a generalized
{\it spin connection} which, in general, depends on position
$X$ in $C$-space.
We formulate the corresponding action principle and derive the
Noether currents in which the charges that generate gauge transformations
are on the same footing as the spin and angular momentum.
We show that the
generalized spin connection contains Yang-Mills fields describing
fundamental interactions, and also the antisymmetric Kalb-Ramond 
\ci{Kalb-Ramond} fields that have an important role in string theory.
So we provide a unified framework for description of Yang-Mills fields,
including Kalb-Ramond fields.

\section{Clifford space as a generalization of spacetime}

Let $V_n$ be a spacetime manifold\footnote{Following the old tradition,
I use the notation $M_n$
for $n$-dimensional (flat) Minkowski space, and $V_n$ for a more
general spacetime manifold, which may have curvature and torsion.
In modern, especially mathematically oriented literature\, \ci{Bruhat,
Rodrigues} a more precise notation is used.}.
For the sake of generality we keep its dimension $n$ arbitrary,
but later we eventually assume the physically observed value $n=4$.
Every point ${\cal P}$ of $V_n$ can be parametrized\footnote{
How precisely this is achieved introducing the charts and atlas has
been explained in many works, therefore I do not repeat it here.}
by arbitrarily chosen coordinates $x^\mu ({\cal P})$. The
subtleties related to the identification of spacetime point were
pointed out by Einstein in his ``hole argument"\,\ci{HoleArgument},
and subsequently
discussed and refined by many authors, of whom let me mention here
DeWitt\,\ci{DeWittFluid} and Rovelli\,\ci{RovelliFluid}. However, what
is clear is
that in spacetime physical {\it events} can happen, and that the
events can be unambiguously identified. If in spacetime there is
a network of events or, in an idealized limit, an infinitely dense
collection of events, e.g., the DeWitt-Rovelli ``reference fluid"
\ci{DeWittFluid,RovelliFluid}, then the spacetime points can be identified.
A quantity of interest is the {\it distance} between two events.
The distance between two infinitely close events in a manifold
endowed with metric is given by
\be
    \dd s^2 = g_{\mu \nu} \dd x^\mu \dd x^\nu
\lbl{2.a1}
\ee
Actualy this is the square of the distance, and $g_{\mu \nu}(x)$
is the metric tensor.

Let us now consider the {\it square root} of the above quadratic
form. Obviously it is $\sqrt{g_{\mu \nu} \dd x^\mu \dd x^\nu}$.
But the latter expression is not linear in $\dd x^\mu$. We
would like to define an object which is {\it linear} in
$\dd x^\mu$ and whose square is eq.\,(\ref{2.a1}). Let such
object be given by the expression\footnote{Here the {\it scalar
components} $\dd x^\mu$ should not be confused with the differential
1-forms.}
\be
    \dd x = \dd x^\mu \gam_\mu
\lbl{2.a2}
\ee
It must satisfy
\be
    \dd x^2 = \gam_\mu \gam_\nu \, \dd x^\mu \dd x^\nu =
    \frac{1}{2} (\gam_\mu \gam_\nu + \gam_\nu \gam_\mu)
    \dd x^\mu \dd x^\nu = g_{\mu \nu} \dd x^\mu \dd x^\nu
\lbl{2.a3}
\ee
from which it follows that
\be
   \gam_\mu \cdot \gam_\nu \equiv {1\oo 2} 
   (\gam_\mu \gam_\nu + \gam_\nu \gam_\mu) = g_{\mu \nu}
\lbl{2.a4}
\ee
This is the defining relation for the generators $\gam_\mu$ of
Clifford algebra. At every point
$x \in V_n$ the objects $\gam_\mu,~\mu=1,2,...,n$, form
a complete set of {\it basis vectors}, and they span a
vector space\footnote{The symbol
$V_n$ used here for a spacetime manifold should not be confused
with a symbol for a vector space. Since the objectives and
the emphasis of research in physics is different from that in
mathematics or mathematical physics, certain discrepancies
of notation are sometimes unavoidable. The symbol $M_4$ is reserved
for Minkowski space (a usual practice in physics), whilst
${\cal M}$ is reserved for the ``membrane space", i.e., the infinite
dimensional space of $r$-loops, considered in ref.\,\ci{PavsicBook}.
The latter space is a generalization of $C$-space, and provides a
possible geometric principle behind the string theory, and is according
to my expectations related to the conjectured $M$-theory which
unifies different string theories.},
called the tangent vector space $T_x (V_n)$.

Basis vectors $\gam_\mu$ occurring in eq.\,(\ref{2.a2}) are
tangent to coordinate curves and they form a {\it coordinate
frame} at $x \in V_n$. More general frames, that are not
tangent to coordinate curves, also exist. Of particular interest
are manifolds such that
at every point $x\in V_n$, or at least at every $x\in {\cal R}$ of
a region ${\cal R} \in V_n$, one can construct an orthonormal
frame.

Eq.\,(\ref{2.a4}) says that the {\it symmetric part} of the
Clifford (or geometric) product $\gam_\mu \gam_\nu$
is the {\it inner product} determining the metric tensor
$g_{\mu \nu}$.

The {\it antisymmetric part} is the {\it wedge product}
determining a bivector representing an oriented area:
\be
    \gam_\mu \wg \gam_\nu \equiv \frac{1}{2}
    (\gam_\mu \gam_\nu - \gam_\nu \gam_\mu)
    \equiv {1\oo 2} [\gam_\mu,\gam_\nu]
\lbl{2..a5}
\ee
This can be continued to the antisymmetric product of
$3,4,...,n$ basis vectors $\gam_\mu$. So the basis
vectors $\gam_\mu$ generate at the point $x \in V_n$
the Clifford algebra ${\cal C \ell}_n$ with the basis
elements
\be
\gam_M \equiv \gam_{\mu_1 \mu_2 ... \mu_r} = \gam_{\mu_1} \wg
\gam_{\mu_2} \wg ... \wg \gam_{\mu_r} \equiv
   \frac{1}{r!} [\gam_{\mu_1},\gam_{\mu_2},...,\gam_{\mu_r}] \; , \quad 
   r=0,1,2,...,n
\lbl{2.a7}
\ee
defined as the antisymmetrized or wedge product.

An important insight that we acquire by involving Clifford
algebra into the game is the following:
\begin{itemize}
   
   \item The ``square root" of the distance quadratic form
   $\dd s^2$ is a {\it vector}.
   
   \item Vectors are Clifford numbers. A complete set of linearly
   independent vectors generate Clifford algebra.
   
   \item Vectors are objects which, like distance, are invariant
   under general coordinate transformations.

\end{itemize}
   
The last point above comes from the fact that under a general
coordinate transformation both the components $\dd x^\mu$
and the basis vectors $\gam_\mu$ change in such manner that
$\dd x = \dd x^\mu \gam_\mu$ remains invariant.

From eqs.\,(\ref{2.a2}),(\ref{2.a4}) we see that in order to
obtain basis vectors $\gam_\mu$ no differentiation is
needed in their definition, if we assume that the metric
tensor $g_{\mu \nu} (x)$ at the point $x \in V_n$ is
already given. By definition $\gam_\mu$ are Clifford
numbers satisfying  relations (\ref{2.a4}). This is a very
direct, and intuitively clear definition of tangent
vectors and the inner product.

In most modern mathematical works on differential geometry,
tangent vectors are identified with differential operator.
Such identification certainly has advantages, otherwise it
would not have been so widely adopted, but it also has
serious drawbacks, as pointed out by Hesteness\, \ci{HestenesVector}.
He wrote: ``It is sufficient to note that if tangent vectors
[defined as differential operators] are not allowed to
generate a geometric algebra at first place, then the
algebra must be artificially imposed on the manifold later on,
because it is absolutely essential for spinors and
quantum mechanics."

It has been shown in many lucid works that Clifford numbers
$\gam_\mu,~\mu=1,2,...,n$, are very useful for description
of geometry and physics, in particular of special and
general relativity. But $\gam_\mu$ alone are not the whole
story. They generate Clifford algebra, spanned by the basis 
(\ref{2.a7}), which brings multivectors ($r$-vectors)
representing oriented areas, volumes, etc., and their superpositions,
called {\it polyvectors},
into the game.
This provides a framework for a theory which goes beyond
the current special or general theory of relativity.

Clifford algebra enables description of, e.g., $r$-dimensional
surfaces associated with extended objects. In order to make
contact with physics, we have to specify which extended
objects we wish to describe. Let us choose to describe
``fundamental" extended objects, such as strings and branes.
Instead of describing an extended object by infinite number
of degrees of freedom, which is one extreme, or only by
the center of mass coordinates, which is another
extreme, we can describe it by a finite number of 
{\it ``polyvector" coordinates}\footnote{In particular cases
those coordinates can be considered as components of a polyvector.
In general this is not so, but for illustrative reasons it is sometimes
convenient to keep the name {\it ``polyvector" coordinates} (see also a
description at the beginning of Sec.\,2.2).}
$x^{\mu_1 ...\mu_r},~
r=0,1,...,n$, which take into account the object's
extension and orientation\,
\ci{Castro,PavsicBook,AuriliaFuzzy,PavsicArena}.

In order to formalize the theory we generalize the notion
of event which, in general, is no longer a {\it point event}
in an $n$-dimensional spacetime. Instead it is an {\it
extended event}, described by a set of ``polyvector"
coordinates $x^{\mu_1 ...\mu_r},~
r=0,1,...,n$. Spacetime is thus replaced
by a larger space, {\it the space of extended events},
called {\it Clifford space} or $C$-space. In the
following we will first review the concept of flat $C$-space
and then generalize it to that of curved $C$-space.

\subsection{Flat Clifford space}

Let us now consider {\it flat} spacetime manifold, i.e.,
the Minkowski space $M_n$. The object $\dd x = \dd x^\mu \gam_\mu$,
defined in eq.\,(\ref{2.a2}), represents a vector joining two
points ${\cal P}$ and ${\cal P}_0$, with coordinates
$x^\mu$ and $x^\mu + \dd x^\mu$, respectively. In flat
manifold this relation can be extended to finite separation of
points:
\be
    x({\cal P}) - x({\cal P}_0) = 
    (x^\mu({\cal P})-x^\mu({\cal P}_0) ) \gam_\mu
\lbl{2.a8}
\ee
Choosing $x^\mu ({\cal P}_0) = 0$, we have
\be
    x({\cal P}) = x^\mu ({\cal P}) \gam_\mu
\lbl{2.a9}
\ee
which is a vector joining the coordinate origin ${\cal P}_0$ and a point
${\cal P}$. In other words, a vector $x$ represents an oriented line,
with one end at the origin ${\cal P}_0$ and the other end at a point
${\cal P}$. If we keep the point ${\cal P}_0$ fixed and ${\cal P}$ variable,
we can say that $x(\cal P)$ denotes a point $\cal P$ of the manifold $M_n$,
or, speaking in physical terms, a {\it point event} in $M_n$. We may
then simplify the notation, and write $x({\cal P}) \equiv x,~
x^\mu ({\cal P}) \equiv x^\mu$, etc.\,.

In eqs.\,(\ref{2.a8}),(\ref{2.a9}), only the basis vectors $\gam_\mu$
occur. Since the latter objects generate Clifford algebra, with the basis
(\ref{2.a7}), one can envisage the following generalization
of eqs.\, (\ref{2.a8}):
\be
    X({\cal E}) - X({\cal E}_0) = \left (x^M ({\cal E})-x^M ({\cal E}_0)
    \right )\gam_M
\lbl{2.a10}
\ee
where ${\cal E}$ generalizes the notion of point events ${\cal P}$.
Choosing $x^M ({\cal E}_0) = 0$, we obtain the corresponding generalization
of eq.\,(\ref{2.a9}):
\be
    X({\cal E}) = x^M ({\cal E}) \gam_M \equiv x^M \gam_M =
    \sigma {\bf 1} + x^\mu \gam_\mu + x^{\mu_1 \mu_2} 
    \gam_{\mu_1 \mu_2} + ... + x^{\mu_1 ... \mu_n} \gam_{\mu_1 ... \mu_n}
\lbl{2.a11}
\ee
where\footnote{If we do not restrict indices to $\mu_1 < \mu_2 < ...$,
then the factors $1/2!,~1/3!,...$ respectively, have to be included in
front of every term in eq.\,(\ref{2.a11}).}
$\mu_1 < \mu_2 < ...$\,. The latter object is a Clifford valued
{\it polyvector},
a superposition of multivectors\footnote{Although Hestenes and others use
the term `multivector' for a generic
Clifford number, we prefer to call it `polyvector', and reserve the name
`multivector' for objects of definite grade. So our nomenclature
is in agreement with the one used in the theory of differential forms,
where 'multivectors' mean objects of definite grade, and not a superposition
of objects with different grades.}
 ($r$-vectors).
It denotes the position of a point in an $2^n$-dimensional Clifford space
($C$-space). The series terminates at a {\it finite} grade,
depending on the dimension $n$. From the point of view of $C$-space
the $r$-vector coordinates
$x^M \equiv x^{\mu_1 ... \mu_r},~r=0,1,2,...,n$,
(called also `polyvector coordinates', since being components of
a polyvector),
denote
a {\it point}. But from the point of view of spacetime they denote
an {\it extended event} ${\cal E}$
associated, e.g., with a closed $(r-1)$-dimensional brane
(called also $(r-1)$-loop)
enclosing an $r$-dimensional
surface, or a superposition of such objects for different grades $r$.
In the case of a closed string (1-loop) embedded in a
target spacetime of $n$-dimensions, one represents the projections
of the closed string (1-loop) onto the embedding spacetime coordinate
planes by the variables $x^{\mu \nu}$. The latter quantities we
call {\it bivector coordinates} of the 1-loop. Similarly for
closed higher dimensional loops\footnote{A more detailed discussion
can be found in ref\,\ci{PavsicArena}.}.

The precise shape of an $(r-1)$-loop is not determined by
the $r$-vector coordinates $x^{\mu_1 ... \mu_r}$. Only
the orientated areas enclosed by the loops are determined.
These are thus {\it collective coordinates} of a loop. They
do not describe all the degrees of freedom of a loop, but
only its collective degrees of freedom---area and 
orientation---common to a family of loops. Namely, in flat $C$-space,
one can obtain the $r$-vector coordinates $x^{\mu_1 ...\mu_r}$
by integrating the oriented $r$-area elements over an oriented
$r$-surface enclosed by the $(r-1)$-loop. For $r$ fixed, there
exist a family of $(r-1)$-loops, all having the same
$x^{\mu_1 ...\mu_r}$.
Therefore,
although the space of loops is infinite dimensional, we obtain a
finite dimensional description, if we identify all loops having the
same coordinates $x^{\mu_1 ... \mu_r}$. This holds for the particular
choice of the $C$-space metric, the diagonal generalized Minkowski
metric, defined in eq.\,(\ref{2.a13}).
 
The space of all possible extended events is $2^n$-dimensional
$C$-space. Since the points of $C$-space are interpreted as
{\it extended events} ``sitting" in spacetime---whose dimension
we eventually fix to the physical value $n=4$---we have that
$C$-space is a physical space, accessible to direct observation.
An observer, according to this theory, can observe 16-dimensional
$C$-space. In this respect the status of $C$-space is
analogous to that of the multidimensional configuration space
of a many particle system. For instance, a system of $N$
particles can be described in terms of a $3N$-dimensional
configuration space, or a $6N$-dimensional phase space.
But basically we still have $3+1$ dimensions, since the
$N$-particle system is ``sitting" in 4-dimensional spacetime.
So, although $C$-space is 16-dimensional, we may still say\textsf{}
that  it describes the physics in 4-dimensional spacetime.
This physics, however, is no longer the ordinary, physics,
but a generalized physics.

In a series of preceding works\,\ci{PavsicBook},\ci{Castro}--\ci{PavsicArena},
\ci{Pezzaglia}--\ci{CastroPavsicReview},\ci{PavsicParis}--\ci{PavsicUnified}
it has been proposed to construct
the extended relativity theory in $C$-space by a natural generalization
of the notion of spacetime interval in Minkowski space to $C$-space:
\be
   \dd S^2 \equiv |\dd X|^2 \equiv \dd X^\ddg * \dd X =
   \dd x^M \dd x^N G_{MN} \equiv \dd x^M \dd x_M
\lbl{2.a12}
\ee
where the metric of $C$-space is given by
\be
   G_{MN} = \gam_M^\ddg * \gam_N
\lbl{2.a13}
\ee
In flat $C$-space, one can choose coordinates $x^M$ such that
$G_{MN} = \eta_{MN} = {\rm diag} (1,1,..., -1,-1,...)$.

The operation $\ddg$ reverses the order of vector:
\be
    (\gam_{\mu_1} \gam_{\mu_2} ... \gam_{\mu_r})^\ddg =
    \gam_{\mu_r} ... \gam_{\mu_2} \gam_{\mu_1}
\lbl{2.a14}
\ee
Indices are lowered and raised by $G_{MN}$ and its inverse
$G^{MN}$. The following relation is satisfied:
 \be
     G^{MJ} G_{JN} = {\delta^M}_N .
\lbl{2.aa9}
\ee

Considering the definition (\ref{2.a13}) for the $C$-space metric, one could
ask why just that definition, which involves reversion, and not a
slightly different definition, e.g., without reversion.
That reversion is necessary for consistency we can demonstrate by the
following example. Let us take a polyvector which has only the 2-vector
component different from zero:
\be
       x^N = (0,0,x^{\alpha \beta},0,0,...,0) .
\lbl{2.aa5a}
\ee
Then the covariant components are
\be
    x_M = G_{MN} x^N = {1\oo 2} G_{M[\alpha \beta]} x^{\alpha \beta}
\lbl{2.aa5b}
\ee
Since the metric $G_{MN}$ is block diagonal, so that $G_{M[\alpha \beta]}$
differs from zero only if $M$ is bivector index, we have
\be
    x_M = x_{\mu \nu} = {1\oo 2} G_{[\mu \nu][\alpha \beta]} x^{\alpha
    \beta} .
\lbl{2.aa5c}
\ee

From the definition (\ref{2.a14}) we find
\be
    G_{[\mu \nu][\alpha \beta]}= (\gam_\mu \wg \gam_\nu)^{\ddagger}
    *(\gam_\alpha \wg \gam_\beta) = (\gam_\nu \wg \gam_\mu)*
    (\gam_\alpha \wg \gam_\beta) = 
    g_{\mu \alpha} g_{\nu \beta} - g_{\mu \beta} g_{\nu \alpha} 
\lbl{2.aa5d}
\ee
where, in particular, $g_{\mu \nu}$ can be equal to Minkowski metric
$\eta_{\mu \nu}$.
Inserting (\ref{2.aa5d}) into (\ref{2.aa5c}) we obtain
\be
    x_{\mu \nu} = {1\oo 2} (g_{\mu \alpha} g_{\nu \beta} - 
    g_{\mu \beta} g_{\nu \alpha}) x^{\alpha \beta} =
    g_{\mu \alpha} g_{\nu \beta} x^{\alpha \beta} 
\lbl{2.aa5e}
\ee
From the fact that the usual metric $g_{\mu \nu}$ lowers the indices
$\mu,~\nu,~\alpha,~\beta,...$, so that
\be
     g_{\mu \alpha} g_{\nu \beta} x^{\alpha \beta} = x_{\mu \nu} 
\lbl{2.aa5f}
\ee
It follows that Eq. (\ref{2.aa5e}) is just an identity.

Had we defined the $C$-space metric without employing the reversion,
then instead of Eq.(\ref{2.aa5d}) and (\ref{2.aa5e}) we would have
$G_{[\mu \nu][\alpha \beta]}=-(g_{\mu \alpha} g_{\nu \beta} - 
g_{\mu \beta} g_{\nu \alpha})$ and $x_{\mu \nu} = - 
g_{\mu \alpha} g_{\nu \beta} x^{\alpha \beta} = - x_{\mu \nu}$,
which is a contradiction\footnote{By an analogous derivation we find
that the relation $x^M = {G^M}_N x^N = {\delta^M}_N x^N$ holds
if ${G^M}_N = (\gam^M)^{\ddagger} * \gam_N$. The definition
${G^M}_N = \gam^M * \gam_N$ leads to the contradictory equation
$x^M = {G^M}_N x^N = - {\delta^M}_N x^N$.}.

Eq.(\ref{2.a12}) is the expression for the {\it line element} in
$C$-space. If $C$-space is generated from the basis vectors
$\gam_\mu$ of spacetime $M_n$ with signature
$(+ - - - - - ...)$, then the signature of $C$-space is
$(+ + + ... - - -...)$, where the number of plus and minus signs is
the same, namely, $2^n/2$. This has some important
consequences that were investigated in ref.\,\ci{PavsicSaasFee}.

We assume that $2^n$-dimensional Clifford space is the
arena in which physics takes place. We can take $n=4$,
so that the spacetime from which we start is just the
4-dimensional Minkowski space $M_4$. The corresponding
Clifford space has then 16 dimensions. In $C$-space the
usual points, lines, surfaces, volumes and 4-volumes are
all described on the same footing and can be transformed
into each other by rotations in $C$-space (called polydimensional
rotations):
\be
    x'^M = {L^M}_N x^N
\lbl{2.aa10}
\ee
subjected to the condition $|\dd X'|^2 = |\dd X|^2$.

\subsection{Curved $C$-space}

So far we have considered {\it flat} $C$-space.
Let us now assume
that in general $C$-space can be curved. As we have the special
relativity in flat spacetime and general relativity
in curved spacetime, so we have now the (extended) special relativity
in flat, and the (extended) general relativity in curved $C$-space. The
notion of $C$-{\it space} generalises that of {\it spacetime}. As a physical
spacetime is dynamical in the sense of being a solution
to the Einstein equations, so according to the proposed theory also a
physical $C$-space is dynamical, being a
solution to the generalized Einstein equations, and can be flat or curved.

Points of a curved $C$-space are described by the coordinates
$x^M \equiv x^{\mu_1 ...\mu_r},~r=0,1,2,...,n$, which we will keep
on calling  polyvector coordinates,
as we did in the case of flat $C$-space. {\it ``Polyvector"  is here
just a name, and it is by no means intended to
assert that, from strict mathematical point of view. $x^M$ are
components of a polyvector}\footnote{However, $x^M$ can be
considered\,\ci{PavsicBook}
as components of a polyvector {\it field} $A^M (X) \gam_M (X)$, such that
in a given coordinate system we have $A^M (X) = x^M$. Then at
every point ${\cal E} \in C$, the object $X({\cal E}) = x^M ({\cal E})
\gam_M ({\cal E})$ is a tangent polyvector. So we have one-to-one
correspondence between the points ${\cal E}$ of the $C$-space manifold
and the polyvector field $X({\cal E}) = x^M ({\cal E}) \gam_M ({\cal E})$
which we will call the coordinate polyvector field. So although
the manifold is curved, every point in it can be described by
a tangent polyvector at that point, whose components are equal to the
coordinates of that point.
We warn the reader not
to confuse the tangent polyvector $X({\cal E})$ at a point ${\cal E}$
with the polyvector joining
the points ${\cal E}_0$ and ${\cal E}$, a concept which is ill
defined in curved manifold $C$.},
joining a fixed point ${\cal E}_0$ (the ``origin") with coordinates
$x^M ({\cal E})_0 =0$ and a variable point ${\cal E}$ with coordinates
$x^M ({\cal E}) \equiv x^M$.
 The name ``polyvector" is used here in
a loose, ``physical" (not mathematical) sense, just to remind us
of the fact that locally, within a sufficiently small region around a chosen
point, a curved $C$-space can be {\it approximated} with a flat $C$-space,
a point of which can be described by a polyvector. Moreover, a curved $C$-space
manifold can be considered as a deformation of flat $C$-space
manifold\footnote{Suppose we have a $C$-space manifold endowed with
the metric ${\bf G} (\epsilon)$ and connection ${\bf D}(\epsilon)$
which depend on a parameter (or a set of
parameters) $\epsilon$, so that for $\epsilon \neq 0$ the manifold
is curved, whilst for $\epsilon \to 0$ it approaches to flat manifold.
Then we say that $\epsilon$ is a deformation parameter (or a set of parameters),
by means of which we deform a flat $C$-space into a curved one.}.
Since the points of flat $C$-space have the interpretation of being 
superpositions of
oriented $(r-1)$-loops (the extended events), the same interpretation is
in place, if we deform the flat $C$-space into a curved
$C$-space.

But in curved $C$-space, unlike in the flat one, coordinates
$x^{\mu_1 ...\mu_r}$ cannot be ``calculated" by performing the
integration of $r$-area elements of an $r$-surface enclosed
by the $(r-1)$-loop. With every event ${\cal E} \in C$ we associate
coordinates,
denoted $x^M, ~M=1,2,...,16$, or equivalently,
$x^{\mu_1 ...\mu_r},~r=0,1,2,3,4$. For this purpose, like in any differential
manifold, we introduce charts that cover different regions
${\cal R}$ of the manifold $C$, and perform the mapping
$x^M:~{\cal E} \in {\cal R} \rightarrow x^M ({\cal E}) \in \mathbb{R}^{16}$,
where $x^M,~M=1,2,...,16$ are real numbers. By assuming that
$C$-space is endowed with metric, we can calculated the distances
between pairs of points ${\cal E},{\cal E}'$. 

However, physically, just the reverse procedure is adopted.
Namely, a smooth,
differential manifold, such as, e.g., $C$-space, is of course
a mathematical idealization. What we have physically, is just
a set of extended events, that can be unambiguously identified,
since they are associated with, e.g,. collisions branes
with $C$-space photon\,\ci{CastroPavsicReview}: the collision
region is an extended region in spacetime, and can be given
arbitrary coordinates $x^M$ (``house numbers"\,\ci{WheelerHouse}).
By measuring the distances between the nearby extended events within
a network of extended events, we can approximately encover
the metric. In order to measure the distance in $C$-space
we suitably generalize the concept of light
clocks\,\ci{Anderson}, which now operate with $C$-space
light rays (based on $C$-space photons). We will postpone a detailed
description to a future paper, since it will require
more practical experience with the applications and consequences
of the special relativity in $C$-space. However, after consulting
the already existing literature \ci{Anderson,PavsicArena,CastroPavsicReview},
an interested reader can straightforwardly work out the basic principles
of $C$-space light clocks and distance measurements. This is beyond the
scope of the present paper which aims at pointing out the potential
of the concept of curved $C$-space for the unification of the
fundamental interactions. Once a sufficient motivation is gained as
a result of such studies, we can go back and work further on what we
have left unsettled.

The distance between two infinitely close extended events is given by
the expression (\ref{2.a12}) (``the line element") in which there
occurs the metric tensor (\ref{2.a13}), given in terms
of the coordinate basis elements $\gam_M$. In flat $C$-space
the coordinate basis elements $\gam_M$ can be chosen so that at {\it every
point} $X({\cal E}) \in C$ they have the same, definite ($r$-vector)
grade, and also so that at {\it every point} $X({\cal E}) \in C$
the metric tensor is diagonal. This is not
the case in curved $C$-space. In curved \cs, the orientation of
a polyvector can
change from point to point. More precisely this means that, if we
perform the parallel transport of a polyvector $A=A^M \gam_M$ from
a point ${\cal E}$ to ${\cal E}'$, the result depends on the path of
transport. If we perform the parallel transport of $A({\cal E})$ from
a point ${\cal E}$ along a closed path back to ${\cal E}$, we obtain
a polyvector $A'({\cal E})$ which differs from $A({\cal E})$. In
particular this means that, if initially we have, e.g., a vector
at ${\cal E}$, then after a round trip parallel transport we can
end up with, e.g., a bivector at ${\cal E}$, or more generally,
with a superposition of bivectors, vectors, 3-vectors, etc.\,.

At every point $X\in C$ the basis elements, that is, the basis
polyvectors, $\gam_M$ span a {\it tangent space} $T_X (C)$ which has
the structure of Clifford algebra ${\cal C\ell}_n$. In a neighbourhood
of a point $X$ the tangent space $T_X (C)$ models the manifold $C$,
i.e., it provides an approximate description of $C$. But in the case
of curved $C$ such description becomes less and less accurate, if we
increase the neighbourhood of $X$.

Basis polyvectors $\gam_M$ are tangent to coordinate curves of $C$-space
and they form a coordinate frame at a point $X\in C$. In general, a set
of $2^n$ linearly independent polyvector fields on a region
${\cal R}$ of $C$-space will be called a frame field.
Of
particular interest are:
\begin{description}
    \item{~(i)} {\it Coordinate frame field} $\{\gam_M\}$. Basis elements
    $\gam_M$, $M= 1,2,...,2^n$ depend on position $X$ in $C$-space.
    The relation (\ref{2.a7}) with wedge product can hold only
    locally at a
    chosen point $X$, but in general it cannot be preserved globally
    at all points $X \in {\cal R}$ of {\it curved} $C$-space.
    The scalar product of two basis elements determines the metric tensor
    of the frame field $\{\gam_M \}$
    according to eq.\,(\ref{2.a13}).
    
    \item{(ii)} {\it Orthonormal frame field} $\{\gam_A\}$.
    Basis elements $\gam_A,~A=1,2,...,2^n$ also depend on $X$. 
    At every point $X$ they are defined as the wedge product   
    and they determine {\it diagonal} metric
\be 
    \gam_A^\ddg * \gam_B = \eta_{AB}
\lbl{2.8} 
\ee
The orthonormal frame field thus consists of the elements
\be
     \lbrace \gam_A \rbrace = \lbrace {\bf 1}, \gam_{a_1}, \gam_{a_1 a_2},...,
     \gam_{a_1 a_2...a_n} \rbrace \; , \quad a_1 < a_2 < ...<a_r \; , \quad
     r=1,2,...,n
\lbl{2.2}
\ee
where
\be
\gam_{a_1 a_2...a_r} = \gam_{a_1} \wg \gam_{a_2} \wg ... \wg
\gam_{a_r} \equiv {1\oo r!} [\gam_{a_1},\gam_{a_2},...,\gam_{a_r}]
\lbl{2.3}
\ee
is the antisymmetrized or wedge product.

\end{description}

The relation between the two sets of basis elements is given in term of
the $C$-space vielbein:
\be
    \gam_M = {e_M}^A \gam_A
\lbl{2.11}
\ee
All quantities in eq.\,(\ref{2.11}) depend on position $X$ in $C$-space.

Explicitly, eq. (\ref{2.11}) reads
\bear
    &&\gam = {e_{\bf o}}^{\underline {\bf o}}\, {\bf 1} + 
    {e_{\bf o}}^{a_1} \gam_{a_1} + {e_{\bf o}}^{a_1 a_2} 
    \gam_{a_1 a_2} +  ...
    + {e_{\bf o}}^{a_1 ... a_n} \gam_{a_1 ... a_n} \nonumber \\
    &&\gam_{\mu_1} = {e_{\mu_1}}^{\underline {\bf o}} \, {\bf 1} + 
    {e_{\mu_1}}^{a_1} \gam_{a_1}
    + {e_{\mu_1}}^{a_1 a_2} \gam_{a_1 a_2} + ... +
    {e_{\mu_1}}^{a_1 ... a_n} \gam_{a_1 ... a_n} \nonumber \\
    &&\gam_{\mu_1 \mu_2} = {e_{\mu_1 \mu_2 }}^{\underline {\bf o}}\, 
    {\bf 1} + 
    {e_{\mu_1 \mu_2}}^{a_1} \gam_{a_1}
    + {e_{\mu_1 \mu_2}}^{a_1 a_2} \gam_{a_1 a_2} + ... +
    {e_{\mu_1 \mu_2}}^{a_1 ... a_n} \gam_{a_1 ... a_n} \nonumber \\
    &&\vdots \nonumber \\
    &&\gam_{\mu_1... \mu_n} = 
    {e_{\mu_1 ... \mu_n }}^{\underline {\bf o}} \,
    {\bf 1} + {e_{\mu_1 ...\mu_n}}^{a_1} \gam_{a_1}
    + {e_{\mu_1 ... \mu_n}}^{a_1 a_2} \gam_{a_1 a_2} + ... +
    {e_{\mu_1 ... \mu_n}}^{a_1 ... a_n} \gam_{a_1 ... a_n}
\lbl{2.12}
\ear
Although here we keep the notation
$\gam_M \equiv \gam_{\mu_1 ... \mu_r}$, the latter
quantities have no definite grade. They had definite grade in flat
$C$-space, where according to eq.\,(\ref{2.a7}) they are equal
to the wedge product of basis vectors. The latter equality is no
longer valid in general, it can hold only locally at a given point
$X$ of curved  $C$-space, but not in its neighbourhood. But for
mnemonic reasons we can retain the notation $\gam_{\mu_1 ...\mu_r}$
and $x^{\mu_1 ...\mu_r}$ even in the neighbourhood of that point.

The relation
(\ref{2.11})
introduced above is a local transformation from
an orthonormal basis $\lbrace \gam_A \rbrace$ to
a coordinate basis $\lbrace \gam_M \rbrace$ of a curved
Clifford space.
A similar relation was considered by Pezzaglia \ci{PezzagliaSpin},
but his interpretation was different, because he imposed
certain constraints in order to satisfy the Clifford algebra
relations. Crawford \ci{Crawford} considered an analogous relation to
that of Pezzaglia, with the difference that the transformation was
from one {\it orthonormal} basis to another {\it orthonormal} basis,
and his quantities depended on spacetime position $x$ only (and not on $X$).
Hence our quantity ${e_M}^A$ is a different object from Pezzaglia's
``geobein" or Crawford's ``drehbein", although the basic idea is
similar, i.e., gauging Clifford algebra.

We have thus a curved {\it Clifford space} ($C$-space). A point of
$C$-space is described by coordinates $x^M$. A coordinate basis at
a point $X$ is $\lbrace
\gam_M|_X \rbrace$, whilst an orthonormal basis is
$\lbrace \gam_A|_X \rbrace $.
The tetrad field is given by the scalar product
${e_M}^A = \gam_M^\ddg * \gam^A$. In particular, 
$\gam_M \equiv \gam_{\mu_1 ... \mu_r}$ at $X\in C$ can be
multivectors of definite grade, i.e., defined as a wedge product
$\gam_{\mu_1} \wedge ... \wedge \gam_{\mu_r}$. But such property can hold
only {\it locally} at  $X$, and cannot be preserved
globally at all points $X$ of our curved Clifford space.

Corresponding to each polyvector field
we define a differential operator, a generalized directional derivative,
which we call {\it derivative} and denote $\p_{\gam_M} \equiv \p_M$, whose
action depends on the
quantity it acts on\footnote{
This operator is a generalization to curved $C$-space of the derivative
$\p_\mu$ which acts in an $n$-dimensional space $V_n$, and was defined by
Hestenes \ci{Hestenes} (who used a different symbol, namely $\Box_\mu$).}:

\ (i) $\p_M$ maps scalars $\phi$ into scalars
\be
    \p_M \phi = {{\p \phi}\oo {\p x^M}}
\lbl{2.15}
\ee
Thus $\p_M$, when acting on scalar fields, is just the ordinary
{\it partial derivative}.

(ii) $\p_M$ maps Clifford numbers into Clifford numbers. In particular,
it maps a coordinate basis Clifford number $\gam_N$ into another Clifford
number which can, of course, be expressed as a linear combination of
$\gam_J$:
\be
     \p_M \gam_N = \Gam_{MN}^J \gam_J
\lbl{2.16}
\ee
The above relation defines the coefficients of connection $\Gam_{MN}^J$
for the coordinate
frame field $\lbrace \gam_M \rbrace$.

An analogous relation we have for the local frame field:
\be
    \p_M \gam_A = - {{\Omega_A}^B}_M \gam_B
\lbl{2.17}
\ee
where ${{\Omega_A}^B}_M$ are coefficients of connection for the local
frame field
$\lbrace \gam_A \rbrace$.

When the derivative $\p_M$ acts on a polyvector valued field
 $A=A^N \gam_N$ we obtain
\be
    \p_M (A^N \gam_N) = \p_M A^N \gam_N + A^N \p_M \gam_N =
    (\p_M A^N + \Gam_{MK}^N A^K) \gam_N \equiv \DD_M A^N \, \gam_N
\lbl{2.17a}
\ee
where $\DD_M A^N \equiv \p_M A^N + \Gam_{MK}^N A^K$ are components of
the covariant derivative in the coordinate basis, i.e., the `covariant
derivative' of the tensor analysis.
Here  $A^N$ are {\it scalar} components of $A$, and $\p_M A^N$ is just
the partial derivative of a scalar field with respect to $X^M$:
\be
    \p_M \equiv \left ( {\p \oo {\p s}},~{\p \oo {\p x^{\mu_1}}},~
    {\p\oo {\p x^{\mu_1 \mu_2}}},..., ~
    {\p \oo {\p x^{\mu_1 ... \mu_n}}} 
    \right )
\lbl{2.18}
\ee

The derivative $\p_M$ behaves as a {\it partial derivative} when acting on
scalar fields, and it defines a {\it connection} when acting on a polyvector
field $\gam_M$ or $\gam_A$. It has turned out very practical
to use the easily writable symbol $\p_M$ which ---when acting on a 
polyvector field--- 
has to be understood in the sense of eq.\,(\ref{2.16})--(\ref{2.17a}).
Especially when doing long
calculation (which is usually the job of
theoretical physicists) it is much easier and quicker to write $\p_M$
than $\Box_M, ~\nabla_M, ~D_{\gam_M}, ~\nabla_{\gam_M}$ which  all are
symbols used in the literature. Also conceptually, it is perhaps
not so wrong to use the same symbols $\p_{\gam_M} \equiv \p_M$, which
so far has been reserved for  the components of
the directional derivative acting on scalar fields, that is, the
partial
derivative of a scalar field. If one tries to perform the partial
derivative of, e.g., a vector field in a curved manifold, one finds that
this cannot be done, unless one specifies how to compare vectors at
different points of the manifold. A common prescription is to perform
a parallel transport of a vector from a point ${\cal P'}$, along a chosen
curve, e.g., along a coordinate line, to a point ${\cal P}$, where
the two vectors can be compared (i.e., substracted).
This is the well
known procedure, therefore we do not attempt to repeat it in detail here.
What we would like to stress is that the
same procedure works in the case of an arbitrary geometric object, including
a scalar. That is, the same definition of the derivative which holds for
an arbitrary geometric object, also holds  for a scalar (in which case
the definition of parallel transport is trivial). Therefore we may retain
the same, unique, symbol $\p_{\gam_M} \equiv \p_M$ for such operator even
when it acts on generic multivector or tensor fields. Here let me remind
the reader that in mathematics there are other cases in which the same
symbol for an operation was retained after extending the set within
which the operation acted. Thus, historically, multiplication was
first defined as acting within the set of (positive) integers. But
subsequently multiplication was generalized to act within increasingly
more general sets of numbers, such as rational numbers, real numbers,
complex numbers, Clifford numbers, etc.. And yet the same symbol,
namely juxtaposition, has been mostly used in algebraic expressions.
We write, e.g, $(a+b)c$ regardless of whether, a,b,c, are integers,
real numbers, complex numbers, Clifford numbers, etc.. Just imagine
how complicated mathematics would have become, if at every generalization
of numbers, we would have invented and used a different symbol for
multiplication.
Perhaps a maximal possible mathematical rigour would indeed require
such notational distinctions, but development of practical calculus and hence
the development of physics and engineering would have been slown down,
if not
completely blocked. After performing many practical calculations\footnote{
See, e.g., how quantum theory in curved space can be elegantly
formulated by using the definition $p = - i \hbar \gam^\mu\p_\mu$
for the momentum operator, which enables, amongst others,
a resolution of the
notorious ordering ambiguities \ci{PavsicOrder}.},
it is
my insight that in the case of Clifford valued fields usage
of a unique symbol $\p_M$ for derivative is a rational choice
which enables a significant simplification and thus renders 
the geometric calculus based on Clifford algebra potentially more
attractive to a wider group of physicists.

There are also other reasons for usage of the unique symbols
$\p_{\gam_M} \equiv \p_M$.
Suppose that we decide to use a symbol such as, e.g., $\DD_{\gam_M}
\equiv \DD_M$ instead
of $\p_{\gam_M}$. The calculation in eq.\,(\ref{2.17a}) would then read
\be
    \DD_{\gam_M} (A^N \gam_N) = \DD_{\gam_M} A^N \gam_N + 
    A^N \DD_{\gam_M} \gam_N =
    (\p_M A^N + \Gam_{MK}^N A^K) \gam_N 
\lbl{2.17b}
\ee
where we have used the property that $\DD_{\gam_M}$ acts on the terms
in a product
according to the Leibnitz rule, and identified $\DD_{\gam_M} A^N = \p_M A^N$.
In order to make a contact with the tensor analysis, which is widely
used in physics, we would need to invent yet another symbol for the
combination $\p_M A^N + \Gam_{MK}^N A^K$. The notation $D_M A^N$ could
not be used for the latter combination, because we already have
$\DD_{\gam_M} A^N \equiv \DD_M A^N = \p_M A^N$. So we would need three
different symbols,
$\p_M,~ \DD_M$, and, say, $\nabla_M$. Moreover, since one says (which is
a common practice in modern differential geometry)
that $\DD_{\gam_M}$ is `covariant derivative', why then 
$(\DD_{\gam_M} A^N) \gam_N= (\p_M A^N) \gam_N$
in eq.\,(\ref{2.17b})
does not transform covariantly (in the sense this word is normally
understood in physics) under general coordinate transformations,
nor does so the term $\Gam_{MK}^N A^K \gam_N$? How can
$\DD_{\gam_M} \gam_N$ be called the ``covariant derivative" of $\gam_N$, if
$\DD_{\gam_M} \gam_N$ does not transform covariantly\footnote{It
is true that the relation $D_M \gam_N = \Gam_{MN}^J \gam_J$ as a whole
is covariant under general coordinate transformations of $x^M$, i.e.,
in new
 coordinates it has the same form as in the old coordinates, but this
does not hold for the terms $D_M \gam_N $ and $\Gam_{MN}^J \gam_J$
taken alone. Namely, under a transformation
 $\gam_N \rightarrow{a_N}^J (X) \gam_J$, we have $\DD_M \gam_N
 \rightarrow {a_M}^K \DD_M ({a_N}^J \gam_J) = {a_M}^K {a_N}^J \DD_M \gam_J
 + ({a_M}^K \p_M {a_N}^J) \gam_J$, which certainly is not a covariant
 transformations property.
}? Only the sum
of the two terms occurring in eq.\,(\ref{2.17b})
transforms covariantly.
If so, then it makes sense not to insist that the operator $\DD_{\gam_M}$
is covariant derivative, but say simply that it is the {\it derivative},
as Hestenes\,\ci{Hestenes} did. When acting on an object (e.g.,
a vector field),
expanded according to $A^N \gam_N$, it automatically acts as a covariant
derivative\footnote{Remember that
in tensor analysis, the ``covariant derivative" when acting according to
the Leibnitz rule to individual terms in a product of tensor field components
retains its
covariant character. For instance, for the product of
vector field components we have
 $\DD_\mu (A^\nu B^\rho)=
(\DD A^\nu) B^\rho + A^\nu (\DD_\mu B^\rho)$, where $\DD_\mu A^\nu$
and $\DD_\mu B^\rho$ transform covariantly under general coordinates
transformations. In the case of a quadratic form, which transforms
as a scalar, we have
$\DD_\mu (A^\alpha B^\beta g_{\alpha \beta}) = (\DD_\mu A^\alpha)
B^\beta + A^\alpha (\DD_\mu B^\beta))g_{\alpha \beta}$, where we have
used $\DD_\mu g_{\alpha \beta} =0$. If we act on the quadratic
form with the partial derivative, we have
$\p_\mu (A^\alpha B^\beta g_{\alpha \beta}) = 
(\p_\mu A^\alpha) B^\beta g_{\alpha \beta} + 
A^\alpha (\p_\mu B^\beta) g_{\alpha \beta}  + A^\alpha B^\beta
\p_\mu g_{\alpha \beta}$ which, after expressing $\p_\mu g_{\alpha \beta}$
by making use of $\DD_\mu g_{\alpha \beta} =0$, turns out to be
the same as $\DD_\mu (A^\alpha B^\beta g_{\alpha \beta})$.
},
but not so if acting on the components $A^N$, or on basis
elements $\gam_N$. If so, it is then better not to use in eq.\,(\ref{2.17b})
the the symbol $\DD_{\gam_M}$ (or whatever other symbol for covariant
derivative),
but use the same symbol as it is used for the partial derivative,
that is, $\p_M$, with understanding that the definition of such operator
is now suitably generalized.

The derivative $\p_M$ is defined with respect to a coordinate frame field
$\lbrace \gam_M \rbrace$ in $C$-space. We can define a more fundamental
derivative $\p$ by
\be
    \p = \gam^M \p_M
\lbl{2.18a}
\ee
This is the {\it gradient} in $C$-space and it generalizes the ordinary
gradient $\gam^\mu \p_\mu$, $\mu = 0,1,2,...,n-1$, discussed by
Hestenes \ci{Hestenes}.

Besides the basis elements $\gam_M$ and $\gam_A$, we can define the
reciprocal elements $\gam^M,~\gam^A$ by the relations
\be
    (\gam^M)^\ddg * \gam_N = {\delta^M}_N \; , \quad
    (\gam^A)^\ddg * \gam_B = {\delta^A}_B
\lbl{2.19}
\ee

\paragraph{Curvature.} We define the curvature of $C$-space in the analogous
way as in the ordinary spacetime, namely by employing the commutator of
the derivatives \ci{Hestenes, PavsicBook, CastroPavsicHigher}.
Using eq.\,(\ref{2.16}) we have
\be
   [\p_M,\p_N] \gam_J = {R_{MNJ}}^K \gam_K
\lbl{2.19a}
\ee
where
\be
    {R_{MNJ}}^K= \p_M \Gam_{NJ}^K - \p_N \Gam_{MJ}^K + 
    \Gam_{NJ}^R \Gam_{MR}^K -
   \Gam_{MJ}^R \Gam_{NR}^K
\lbl{2.20}
\ee
is the curvature of $C$-space. Using (\ref{2.19a}) we can express the
curvature according to
\be
    (\gam^K)^\ddg * ([\p_M,\p_N] \gam_J) = {R_{MNJ}}^K
\lbl{2.21}
\ee

An analogous relation we have if the commutator of the derivatives operates
on a local basis elements and use eq.\,(\ref{2.17}):
\be
   [\p_M,\p_N] \gam_A = {R_{MNA}}^B \gam_B
\lbl{2.22}
\ee
where
\be
    {R_{MNA}}^B= -(\p_M {{\Omega_A}^B}_N - \p_N {{\Omega_A}^B}_M
    + {{\Omega_A}^C}_N {{\Omega_C}^B}_M - {{\Omega_A}^C}_M {{\Omega_C}^B}_N)
\lbl{2.23}
\ee

In view of eq.\,(\ref{2.11})  an arbitrary polyvector can be expanded
according to
\be
     A = A^M \gam_M = A^M {e^A}_M \gam_A
\lbl{2.24}
\ee
In particular we have
\be
    \p_N \gam_M = \Gam_{MN}^J \gam_J = \p_N ({e^A}_M \gam_A) =
    \p_N {e^A}_M \gam_A + {e^A}_M \p_N \gam_A
\lbl{2.25}
\ee
Using (\ref{2.16}),(\ref{2.17}), we obtain from (\ref{2.25}) the following
relation which involves $C$-space vielbein field and the
coefficients of connection for the frame filed $\lbrace \gam_M \rbrace$
and $\lbrace \gam_A \rbrace$, respectively:
\be
    \p_N {e^C}_M - \Gam_{NM}^J {e^C}_J - {e^A}_M {{\Omega_A}^C}_N = 0
\lbl{2.26}
\ee
This is a generalization of the well known relation in an ordinary curved
spacetime. In eq.\,(\ref{2.26}) ${{\Omega_A}^C}_N$ extends the notion
of the spin connection coefficients ${{\omega_a}^c}_\nu$.

From (\ref{2.26}) we obtain
\be
    \p_M {e^C}_N - \p_N {e^C}_M + {e^A}_M {{\Omega_A}^C}_N 
    - {e^A}_N {{\Omega_A}^C}_M = {T_{MN}}^J {e^C}_J
\lbl{2.27}
\ee
where
\be   
      {T_{MN}}^J  = \Gam_{MN}^J - \Gam_{NM}^J
\lbl{2.28}
\ee
is the $C$-space {\it torsion}.

In general, the torsion is different from zero. In particular, when torsion
vanishes, we find the following expression for the connection coefficients:
\be
    \Omega_{BCM} = {1\oo 2} {e^A}_M (\Delta_{[AB]C} - \Delta_{[BC]A} +
    \Delta_{[CA]B})
\lbl{2.29}
\ee
where
\be
      \Delta_{[AB]C} \equiv {e_A}^M {e_B}^N (\p_M e_{NC} - \p_N e_{MC})
\lbl{2.30}
\ee
generalizes the notion of the Ricci rotation coefficients.

\section{On the general relativity in $C$-space}

The basic idea of the novel theory that we are exploring here is that the
concept of spacetime should be replaced by that of Clifford space.
Although the name ``Clifford space" could sound very mathematical and
thus not much related to physics, just the contrary is true. Clifford
space ($C$-space) is the very space in which physics takes place. By
considering so far only spacetime we have omitted a big portion of
a very relevant physics which has been sitting just around the corner.
Namely, as we have shown in this and previous works, spacetime is just
the start. From its basis we can build a larger space,
which is Clifford space.
And the latter space, according to the view held in this and other
papers \ci{Pezzaglia}--\ci{CastroPavsicReview},\ci{PavsicBook},
\ci{Castro}--\ci{PavsicArena}, 
is just as physical as the spacetime of general relativity.
Since $C$-space has more than four dimensions (namely 16, if built on
4-dimensional spacetime), it can serve as a realization of the Kaluza-Klein
theory.

We have thus a 16-dimensional, ``physical", continuum, whose points are
described by coordinates $x^M = (s,x^{\mu_1},x^{\mu_a \mu_2},...,
x^{\mu_1 ... \mu_n})$. Using a frame field, say, a coordinate frame field
$\lbrace \gam_M \rbrace$, the metric is given by $G_{MN} =
\gam_M^\ddg * \gam_N$. The basis elements $\gam_M = {e^A}_M \gam_A$,
$M=({\bf o},[\mu_1],[\mu_1 \mu_2],...,
[\mu_1 ... \mu_n]),~ \mu_1<\mu_2<...<\mu_r$, or equivalently,
the metric $G_{MN}$, are considered as dynamical variables of the theory.
From the curvature, defined in eqs.\,(\ref{2.19})--(\ref{2.21}) we can
form a kinetic term for $\gam_M$, or equivalently, $G_{MN}$.

In addition, we also have sources. The first straightforward possibility
is to introduce a single parameter $\tau$ and consider a mapping
\be
    \tau \rightarrow x^M = X^M (\tau)
\lbl{3.1}
\ee
where $X^M (\tau)$ are 16 embedding functions that
describe a worldline in $C$-space. From the point of view of
$C$-space, $X^M (\tau)$ describe a worldline of a ``point
particle": at every value of $\tau$ we have a {\it point} in
$C$-space. But from the perspective of the underlying
4-dimensional spacetime, $X^M (\tau)$ describe an extended
object, sampled by the center of mass coordinates
$X^\mu (\tau)$ and the coordinates
$X^{\mu_1 \mu_2}(\tau),..., X^{\mu_1 \mu_2 \mu_3 \mu_4} (\tau)$.
They are a generalization of the center of mass coordinates in the sense
that they provide information about the object's 2-vector, 3-vector, and
4-vector extension and orientation. For instance, in the case of a closed
string we have a 2-dimensional
surface enclosed by a 1-dimensional line. Integrating over the oriented area
elements, we obtain a finite effective oriented area given in terms
of bivector coordinates $X^{\mu \nu}$. Such bivector coordinates provide
an approximate description of a closed string; they do not provide a complete
description of the string, but nevertheless, they provide a better
approximation, than the mere center of mass coordinates\,\ci{PavsicArena}.

As already discussed in Sec.\,2.2,
the integration described above and in ref.\,\ci{PavsicArena} makes sense
in flat spacetime. Then one can choose as simple a metric tensor
as possible, i.e., the diagonal metric tensor, and perform the integration
over an $r$-surface enclose by an  $(r-1)$-loop which gives the
polyvector coordinates $x^{\mu_1 ... \mu_r}$. In curved spacetime this
cannot be done. One has to consider an extended object, e.g., a brane,
as associated with an {\it extended event} ${\cal E}$, to which we can
assign arbitrary coordinates (``house numbers" according to Wheeler),
i.e., perform a mapping $x^M: {\cal E} \rightarrow \mathbb{R}^{16}$.
Suppose
now that there is a network of extended events, in an idealized limit, an
infinitely dense collection of extended events forming altogether
a 16-dimensional manifold C. If we measure the distances
between the neighbouring extended events, we obtain the metric tensor.
This is how we can arrive, starting from physics, at the abstract
notion of a $C$-space manifold, endowed with metric (see also Sec.\,2.2).

Let us assume that the classical action contains a term which
describes the ``point particle" in $C$-space and a kinetic term
which describes the dynamics of the $C$-space itself:
\be
    I[X^M,G_{MN}] = \int \dd \tau \, ({\dot X}^M {\dot X}^N G_{MN})^{1/2}
    + {\kappa \oo {16 \pi}} \int [\dd x] R
\lbl{3.2}
\ee
Here ${\dot X}^M \equiv \dd X^M/\dd \tau$, $\kappa$ a constant,
$[\dd x]\equiv \dd^{16} x$ the measure on $C$-space, and $R={R_{MNJ}}^N$
the curvature scalar
(see eqs.\,(\ref{2.19a})--(\ref{2.23}))
of $C$-space, analogous to the curvature scalar of the ordinary general
relativity. The action is invariant under local (pseudo) rotations
in tangent space $T_X (C)$, and
under general coordinate transformations in $C$-space.

Variation of the action (\ref{3.2}) with respect to $X^M$ gives
the geodetic equation in $C$-space:
\be
{1\oo {\sqrt{{\dot X}^2}}} \,
     {\dd \oo {\dd \tau}} \left ( {{\dot X}^M\oo \sqrt{{\dot X}^2} } 
     \right )
     + \Gam_{JK}^M {{{\dot X}^J {\dot X}^K}\oo {\dot X}^2 } = 0
\lbl{3.3}
\ee

Varying (\ref{3.2}) with respect to $G_{MN}$ gives the $C$-space Einstein's
equations
\be
    R^{MN} - {1\oo 2} G^{MN} R = 8 \pi \kappa \int \dd \tau \,
    \delta^{(C)} (x - X(\tau)) {\dot X}^M {\dot X}^N
\lbl{3.4}
\ee
where $\delta^{(C)}$ is the $\delta$-function in $C$-space.

When looking from the 4-dimensional spacetime, the equation of geodesic
(\ref{3.3}) contains besides the usual gravitation also other
interactions. They are encoded in the metric components $G_{MN}$ of
$C$-space. Gravity is related to the components $G_{\mu \nu},~\mu,\nu
=0,1,2,3$, while the gauge fields due to other interactions
are related to the components 
$G_{\mu {\bar M}}$,
where the index ${\bar M} \neq \nu$ assumes 12 possible values, excluding
the four values $\nu = 0,1,2,3$. In addition, there are also
interactions due to the components $G_{{\bar M} {\bar N}}$, but they
have not the property of the ordinary Yang-Mills fields.

If we now consider the known fundamental interactions of the standard model
we see that besides gravity we have 1 photon described by the abelian
gauge field $A_\mu$, 3 weak gauge bosons described by gauge fields
$W_\mu^a$, $a=1,2,3$,  and 8 gluons described $A_\mu^c$, $c=1,2,...,8$.
Altogether there are 12 gauge fields.

Interestingly, the number of mixed components $G_{\mu {\bar M}} =
(G_{\mu [{\bf o}]},~ G_{\mu[\alpha \beta]},~G_{\mu[\alpha \beta \rho]},~
G_{\mu [\alpha \beta \rho \sigma]})$ of the $C$-space metric tensor
$G_{MN}$ coincides with the number of gauge fields in the standard
model\footnote{The numbers of the independent indices $[{\bf o}],~
[\alpha \beta],~[\alpha \beta \rho],~[\alpha \beta \rho \sigma]$ are
respectively 1,6,4,1 which sums to 12.}. For fixed $\mu$, there are 12
mixed components of $G_{\mu {\bar M}}$ and 12 gauge fields $A_\mu,~
W_\mu^a, A_\mu^c$. This coincidence is fascinating and it may indicate
that the known interactions are incorporated in curved Clifford space.

Good features of $C$-space are the following:
\begin{description}
 
  \item[\ \ (i)] We do not need to introduce extra dimensions of spacetime.
  We stay with 4-dimensional spacetime manifold $V_4$, and yet we can proceed
  \` a la Kaluza-Klein. The extra degrees of freedom are in $C$-space,
  which is the space of extended events associated with extended
  objects residing in spacetime manifold $V_4$. 
  
  \item[\ (ii)] We do not need to compactify the extra ``dimensions". The
  extra dimensions of $C$-space, namely $s,~x^{\mu \nu},~
  x^{\mu \nu \rho},~x^{\mu \nu \rho \sigma}$ are not just like the ordinary
  dimensions of spacetime considered in the usual Kaluza-Klein theories.
  The coordinates $x^{\mu \nu},~x^{\mu \nu \rho},~x^{\mu \nu \rho \sigma}$
  are related to oriented $r$-surfaces, $r=2,3,4$, by which we sample
  extended objects. Those degrees of freedom are in principle not
  hidden from our direct observation, therefore we do not need to
  compactify such ``internal" space.
  
  \item[(iii)] The number of the mixed metric components $G_{\mu {\bar M}}$
  (for fixed $\mu$) is 12, precisely the same as the number of gauge fields
  in the standard model.

\end{description}

We will not go in further details, since they have already been
written down in Kaluza-Klein theories, although with a different
interpretation of the extra dimensions.
However, one cannot expect to obtain a realistic theory, with correct
coupling constants, within the realm of a classical theory.

\section{The generalized Dirac equation in curved $C$-space}

\subsection{Spinors as members of left ideals of Clifford algebra}

How precisely the curved $C$-space is related to Yang-Mills gauge
fields can be
demonstrated by considering a generalization of the Dirac equation to
curved $C$-space.

Let $\Phi (X)$ be a polyvector valued field over coordinate polyvector
field $X= x^M \gam_M$:
\be
    \Phi = \phi^A \gam_A
\lbl{4.1}
\ee
where the basis elements $\gam_A,~A=1,2,...,16$, form an orthonormal frame
field on a region
${\cal R}$ of $C$-space\footnote{Of particular interest are such $C$-spaces
in which there exist global orthonormal frames fields.}
(see eq.(\ref{2.2})) and $\phi^A$  are the projections (components) of
$\Phi$ onto $\gam_A$. We will suppose that in
general $\phi^A$ are complex-valued scalar quantities.

We interpret the imaginary unit $i$ in the way that is usual in quantum
theory, namely that $i$ lies outside the Clifford algebra of spacetime and
hence commutes with all $\gam_M$. This is different from the point of
view hold by many researchers of the geometric calculus based on Clifford
algebra (see, e.g., \ci{Hestenes, Lounesto}). They insist that $i$ has to
be defined geometrically, so it must be one of the elements of the set
$\lbrace \gam_A \rbrace$, such that its square equals $-1$. An alternative
interpretation, also often assumed, is that $i$ is the pseudoscalar 
unit of a 
higher dimensional space. For instance, if our spacetime
is assumed to be 4-dimensional, then $i$ is the pseudoscalar unit of a
5-dimensional space. The problem then arises about a physical interpretation
of the extra dimension. This is not the case that we adopt. Instead we
adopt the view, first proposed in \ci{PavsicBook}, that $i$ is the bivector
of the 2-dimensional {\it phase space} $P_2$, spanned by $e_q,~e_p$, so that
$Q\in P_2$ is equal to $Q=e_q e_q + p e_p,~ e_q Q = q + i p, ~ i=e_q e_p$.
So our $i$ is also defined geometrically, but the space we employ
differs from
the spaces usually considered in defining $i$. Taking into account that
there are four spacetime dimensions, the total phase space is thus the
direct product $M_4 \times P_2 = P_8$, so that any element $Q \in P_8$ is
equal to $Q= x^\mu e_\mu e_p + p^\mu e_\mu e_p,~~e_q Q = (x^\mu + i p^\mu)
e_\mu$. This can then be generalized to Clifford space by replacing
$x^\mu,p^\mu$ by the corresponding Clifford space variables $x^M, p^M$.
In a classical theory, we can just consider $x^\mu$ only 
(or $x^M$ only), and
forget about $p^\mu$ ($p^M$), since $x^\mu$ and $p_\mu$ are
independent. In
quantum theory, $x^\mu$ and $p_\mu$ ($x^M$ and $p_M$) are complementary
variables, therefore we cannot formulate a theory without at least implicitly
involving the presence of momenta $p_\mu$ ($p_M$). Consequently, wave
functions are in general complex valued. Hence the occurrence of $i$ in
quantum mechanics is not perplexing, it arises from phase space.
We adopt here the conventional interpretation of quantum mechanics:
no hidden variables, B\" ohmian potential, etc., just the Born statistical
interpretation and Bohr-Von Neumann projection postulate. The formalism
described here works for the Everett interpretation as well.

Instead of the basis\footnote{We will use `basis'
and `frame' as synonims. In order to simplify notation and
wording, we will be sloppy in distinguishing objects from the corresponding
fields, e.g., (poly)vectors from (poly)vector fields,
frames from frame fields, (generalized) spinors from (generalized)
spinor fields etc. From the context it should not be difficult to
understand when the talk is about an object taken at a point $X \in C$,
and when about an infinite dimensional object, a field. Thus
$\gam_A$, $A=1,2,...,16$, can mean either orthonormal polyvectors
taken at a point $X \in C$, or polyvector fields, depending on the
context. 
Similarly $\lbrace \gam_A \rbrace$ can mean either an orthonormal
frame (basis), or orthonormal frame field. Thus we avoid the
notation $\lbrace \gam_A|_X \rbrace$, which is unnecessarily clumpsy
within a physically oriented paper. The situation, of course is different
in mathematically oriented papers, where maximal possible rigour and
very precise notation are requested.}
 $\lbrace  \gam_A \rbrace$ one can consider another
basis, which is obtained after multiplying $\gam_A$ by 4 independent
primitive idempotents \ci{Teitler}
\be
    P_i = {1\oo 4} ({\bf 1} + a_i \gam_A + b_i \gam_B + 
    c_i \gam_C) \; , \quad
   i=1,2,3,4
\lbl{4.2}
\ee
such that
\be
    P_i = {1\oo 4} ({\bf 1} +a_i \gam_A)({\bf 1}+b_i \gam_B) \; , 
    \quad \gam_A \gam_B
   = \gam_C \; , \quad c_i = a_i b_i
\lbl{4.3}
\ee
Here $a_i,~b_i,~c_i$ are complex numbers chosen
so that $P_i^2 = P_i$. For explicit and systematic
construction see \ci{Teitler, MankocIdeal}.

By means of $P_i$ we can form minimal ideals of Clifford algebra. A
basis of a left (right) minimal ideal is obtained by taking one of
$P_i$ and multiply it from the left (right) with all 16 elements
$\gam_A$ of the algebra:
\be
   \gam_A P_i \in {\cal I}_i^L \; , \qquad P_i \gam_A \in {\cal I}_i^R
\lbl{4.4}
\ee
Here ${\cal I}_i^L$ and ${\cal I}_i^R$, $i=1,2,3,4$ are four independent
minimal left and right ideals, respectively. For a fixed $i$ there are
16 elements $\gam_A P_i$ or $P_i \gam_A$, but only four amongst
them are different, the remaining elements are just
repetition---apart from constant factors---of those four different elements.

Let us denote those different elements $\xi_{\alpha i}$, $\alpha=
1,2,3,4$. They form a basis of the $i$-th left ideal.

As an illustration let us provide an example. Let
\bear
    P_1 = {1\oo 4} (1 + \gam_0 + i \gam_{12} + i \gam_{012}) \lbl{4.7a}\\
    P_2 = {1\oo 4} (1 + \gam_0 - i \gam_{12} - i \gam_{012}) \lbl{4.7b}\\
    P_3 = {1\oo 4} (1 - \gam_0 + i \gam_{12} - i \gam_{012}) \lbl{4.7c}\\
    P_4 = {1\oo 4} (1 - \gam_0 - i \gam_{12} + i \gam_{012}) \lbl{4.7d}
\ear
In short,
\be
    P_i ={1\oo 4} (1 \pm \gam_0)(1\pm i \gam_{12})
\lbl{4.8}
\ee
where 4 different choices of sign give 4 different idempotents $P_i$.

The basis of the first left ideal is
\bear
   \xi_{11} &=& P_1 = {1\oo 4} (1 + \gam_0 + i \gam_{12} + i \gam_{012})
   \nonumber \\
   \xi_{21} &=& - \gam_{13} P_1 = {1\oo 4}(-\gam_{13} - \gam_{013} + i \gam_{23}
   + i \gam_{023}) \nonumber \\
   \xi_{31} &=& - \gam_3 P_1 = {1 \oo 4} (-\gam_3 + \gam_{03} - i \gam_{123}
   + i \gam_{0123}) \nonumber \\
   \xi_{41} &=& - \gam_1 P_1 = {1\oo 4}(-\gam_1 + \gam_{01} + i \gam_2
   - i \gam_{02})
\lbl{4.9}
\ear
All sixteen basis elements $\gam_A = ({\bf 1},\gam_{a_1},\gam_{a_1 a_2},
\gam_{a_1 a_2 a_3}, \gam_{a_1 a_2 a_3 a_4}),~a_1<a_2<...<a_r,~r=0,1,2,3,4$,
take place in equation (\ref{4.9}) defining $\xi_{\alpha 1}$. For
$i=2,3,4$ we have  expressions analogous to (\ref{4.9}), with suitably
changed signs and order of indices $\alpha = 1,2,3,4$.
Altogether there are 16 basis elements $\xi_{\alpha i}$, $i=1,2,3,4$.
The basis $\lbrace \xi_{\alpha i} \rbrace$ is complete. Every Clifford
number can be expanded either in terms of $\gam_A$ or in terms of
$\xi_{\alpha i} =
(\xi_{\alpha 1},~\xi_{\alpha 2},~\xi_{\alpha 3},~\xi_{\alpha 4})$:
\be
   \Phi = \phi^A \gam_A = \Psi = \psi^{\alpha i} \xi_{\alpha i} =
\psi^{\tilde A} \xi_{\tilde A}
\lbl{4.10}
\ee
In the last step we introduced a single spinor index ${\tilde A}$
which runs over all 16 basis elements that span 4 independent
left minimal ideals so that $\xi_{\tl A} = 
(\xi_{\alpha 1},~\xi_{\alpha 2},~\xi_{\alpha 3},~\xi_{\alpha 4})$.   
Explicitly, eq. (\ref{4.10}) reads
\be
   \Psi = \psi^{\tilde A} \xi_{\tilde A} = \psi^{\alpha 1} \xi_{\alpha 1}
  + \psi^{\alpha 2} \xi_{\alpha 2} + \psi^{\alpha 3} \xi_{\alpha 3} +
\psi^{\alpha 4} \xi_{\alpha 4}
\lbl{4.11}
\ee
Eq.(\ref{4.11}) or (\ref{4.10}) represents a direct sum of four independent
4-component spinors, each in a different left ideal ${\cal I}_i^L$.

The spinor basis elements $\xi_{\tl A}$ are related to the Clifford
algebra basis elements $\gam_A$ according to
\be
    \xi_{\tl A} = {H^B}_{\tl A} \, \gam_B
\lbl{4.12}
\ee
where ${H^B}_{\tl A} $ is a matrix that can be read from eqs.\,(\ref{4.9}),
and the analogous equation for the remaining three left ideals.
An element $\xi_{\tl A}$ of the spinor basis is a
superposition of the elements $\gamma_A$. The coefficients
${H^B}_{\tl A}$ of
the superposition contain real or imaginary numbers. 

The metric of the local {\it flat} tangent space is given by the scalar
product
\be
    \gam_A^\ddg * \gam_B = \langle \gam_A^\ddg \gam_B \rangle_0 = 
    G_{AB} {\bf 1}
\lbl{4.13}
\ee
which involves the reversion operation $\ddg$ (it reverses the order of
vectors entering $\gam_A \equiv \gam_{a_1 ... a_r}$).

Analogously we can define the metric in terms of the spinor basis elements
$\xi_{\tl A}$:
\be
     \xi_{\tl A}^\ddg * \xi_{\tl B} = \langle \xi_{\tl A}^\ddg \xi_{\tl B}
     \rangle_0 = {1\oo n} Z_{{\tl A}{\tl B}} {\bf 1}
\lbl{4.14}
\ee
${\bf 1}$ being the unit element of the Clifford algebra. Reversion
here acts on {\it all} basis elements entering the definition of
$\xi_{\tl A}$ (eq.\,(\ref{4.12})): not only on $\gam_A$, but also on
the imaginary unit $i \equiv e_q e_P$, the bivector of phase space,
(occurring in ${H^B}_{\tl A}$, e.g.,  in eq.\, (\ref{4.9})), so that
$i^\ddg = -i$. This is just the complex conjugation $i^* = -i$.

The occurrence of $n= {\delta_\mu}^\nu = 4$ (i.e., the dimension of the
spacetime from which we generate Clifford algebra) comes from a choice
of normalization constant in the definition of $\xi_{\tl A}$ (e.g.,
in eq.\,(\ref{4.9})). Let us define an operation $\langle ~\rangle_S$,
distinct from $\langle ~\rangle_0$ with the properties:

a) for the unit element ${\bf 1}$
and an arbitrary Clifford number $A$ we have
\bear
      &\langle {\bf 1} \rangle_S = n&
 \lbl{4.14a}
\\
      &\langle A \rangle_S = n \langle A \rangle_0&
 \lbl{4.15}
\ear

b) for the product of Clifford numbers we have the cyclic behavior:
\be
  \langle A B \rangle_S = \langle B A \rangle_S \; , \quad
  \langle A B C \rangle_S = \langle C B A \rangle_S \; , \quad
  \langle A_1 A_2 ... A_k \rangle_S = \langle A_k A_{k-1}... A_1 \rangle_S
\lbl{4.14b}
\ee
Using the operation $\langle A \rangle_S$ we find
\be
   \langle \xi_{\tl A}^\ddg \xi_{\tl B} \rangle_S = Z_{{\tl A}{\tl B}}
\lbl{4.15a}
\ee

Introducing the inverse matrix $Z^{{\tl A}{\tl B}}$ according to
\be
    Z^{{\tl A}{\tl C}} Z_{{\tl C}{\tl B}} = {\delta^{\tl A}}_{\tl B}
\lbl{4.16}
\ee
we have $\xi_{\tl A} = Z^{{\tl A}{\tl C}} \xi_{\tl C}$,~ 
${\xi_{\tl A}}^\ddg = Z^{{\tl A}{\tl C}} \xi_{\tl C}^\ddg$ and
\be
     \langle {\xi^{\tl A}}^\ddg \xi_{\tl B} \rangle_S = 
     {\delta^{\tl A}}_{\tl B}
\lbl{4.17}
\ee
With respect to the operation $\langle ~~ \rangle_S$ the spinor basis
elements $\xi_{\tl A}$ are orthonormal in the above sense.

Relations (\ref{4.15a}),(\ref{4.17}) can be directly verified from
eq.\,(\ref{4.9}) and similar relations for $\xi_{\alpha i}$, $i=2,3,4$.

Matrix elements of an arbitrary Clifford number $A$ in the (generalized)
spinor basis can be calculated according to
\be
    \langle \xi_{\tl A}^\ddg A \xi_{\tl B} \rangle_S 
    \equiv A_{{\tl A}{\tl B}}
    \; , \quad 
\langle {\xi^{\tl A}}^\ddg A \xi_{\tl B} \rangle_S 
\equiv {A^{\tl A}}_{\tl B}
\lbl{4.18}
\ee
For $A$ we may take the basis elements $\gam_A$, and in particular the
generators $\gam_a$, $a=0,1,2,3$. So we obtain the Dirac matrices as a
particular case:
\be
    \langle \xi_{\alpha}^\ddg \gam_a \xi_{\beta} \rangle_S
     = (\gam_a)_{\alpha
    \beta} \quad {\rm and} \quad
    \langle {\xi^{\alpha}}^\ddg \gam_a \xi_{\beta} \rangle_S = 
    {(\gam_a)^\alpha}_\beta 
\lbl{4.19}
\ee
where $\alpha,~\beta$ are the 4-spinor indices, and $\xi_{\alpha} \equiv
\xi_{\alpha 1},~\xi_\beta \equiv \xi_{\beta 1}$.

The quadratic form of a polyvector $\Phi = \phi^A \gam_A = 
\psi^{\tl A} \xi_{\tl A}$ is
\be
   \Phi^\ddg * \Phi = \langle \Phi^\ddg \Phi \rangle_0 =
   \langle \phi^{*A} \gam_A^\ddg \gam_B \phi^B \rangle_0 = 
   \phi^{*A} G_{AB} \phi^B {\bf 1}
\lbl{4.20}
\ee
or
\be
   \langle \Psi^\ddg \Psi \rangle_S = \langle \psi^{*{\tl A}} \psi^{\tl B}
   \xi_{\tl A}^\ddg \xi_{\tl B} \rangle_S = 
   \psi^{*{\tl A}} Z_{{\tl A}{\tl B}} \psi^{\tl B}
\lbl{4.21}
\ee
In eq.\,(\ref{2.20}) we take the scalar part of the expression, whilst in
eq.\,(\ref{2.21}) we perform the operation (\ref{4.15}). The latter operation
is equivalent to taking the trace of the matrix representing the
Clifford number.

Explicitly, for the basis (\ref{4.9}) of the first left ideal we have\footnote{
We omit the index $i$, denoting the ideal, since the same relation holds
for each ideal.}
\be
    \langle \xi_{\alpha}^\ddg \xi_{\beta} \rangle_S = z_{\alpha \beta}
    = \begin{pmatrix}  1 & 0 & 0 & 0 \\
                 0 & 1 & 0 & 0 \\
                 0 & 0 & -1 & 0 \\
                 0 & 0 & 0 & -1 \end{pmatrix}
\lbl{4.22}
\ee
and
\be
     \langle {\xi^{\alpha}}^\ddg \xi_{\beta} \rangle_S = {\delta^\alpha}_\beta
\lbl{4.22a}
\ee

For the basis $\xi_{\tl A} = \xi_{\alpha i}$, spanning all four ideals,
we have
\be
    \langle \xi_{\tl A}^\ddg \xi_{\tl B} \rangle_S = Z_{{\tl A}{\tl B}}
    = \begin{pmatrix}  1 & 0 & 0 & 0 \\
                 0 & 1 & 0 & 0 \\
                 0 & 0 & 1 & 0 \\
                 0 & 0 & 0 & 1 \end{pmatrix} \otimes
      \begin{pmatrix}  1 & 0 & 0 & 0 \\
                 0 & 1 & 0 & 0 \\
                 0 & 0 & -1 & 0 \\
                 0 & 0 & 0 & -1 \end{pmatrix}  
\lbl{4.23}
\ee

We can calculate the matrix elements of the basis Clifford numbers $\gam_A$
as follows:
\be
     \langle {\xi^{\tl A}}^\ddg \gam_A \xi_{\tl B} \rangle_S =
     {(\gam_A)^{\tl A}}_{\tl B} = {\delta^i}_j \otimes {(\gam_A)^\alpha}_\beta
\lbl{4.24}
\ee
For instance, the matrix elements of basis vectors $\gam_a$ are given by
\be
\langle {\xi^{\tl A}}^\ddg \gam_a \xi_{\{\tl B} \rangle_S =
     {(\gam_a)^{\tl A}}_{\tl B} = {\delta^i}_j \otimes {(\gam_a)^\alpha}_\beta
\lbl{4.25}
\ee
where
\be
     {(\gam_a)^\alpha}_\beta   = 
     \langle {\xi^{\alpha}}^\ddg   \gam_a \xi_\beta \rangle_S
\lbl{4.26}
\ee
are the usual Dirac matrices. Using the spinor basis (\ref{4.9}) we
obtain from (\ref{4.26}) just the matrices in the Dirac representation.

Notice that we define the Dirac matrices by taking one contravariant and
one covariant spinor index. Such convention is embraced by the relations
(\ref{4.22}),(\ref{4.22a}), according to which the unit matrix ${\bf 1}$
is replaced by ${\delta^\alpha}_\beta$. The Clifford algebra relations
\be
    \gam_a \gam_b + \gam_b \gam_a = 2 \eta_{ab} {\bf 1}
\lbl{4.27}
\ee
can be written in the matrix form according to
\be
  \langle {\xi^{\alpha}}^\ddg (\gam_a \gam_b + \gam_b \gam_a) \xi_\beta
  \rangle_S = 2 \eta_{ab} \langle {\xi^{\alpha}}^\ddg {\bf 1} \xi_\beta
  \rangle_S
\lbl{4.28}
\ee
Let us now take into account the relation
\be
    \xi_\rho {\xi^\rho}^\ddg = {\bf 1}
\lbl{4.29}
\ee
which can be directly calculated from eq.\,(\ref{4.9}), and insert it into
eq.\,(\ref{4.28}). We obtain
\be
  \langle {\xi^{\alpha}}^\ddg \gam_a \xi_\rho {\xi^\rho}^\ddg \gam_a \xi_\beta
  \rangle_S + (a \stackrel{\leftarrow}{\rightarrow} b) = 2 \eta_{ab}
  {\delta^\alpha}_\beta
\lbl{4.28a}
\ee
which can be written as
\be
   {(\gam_a)^\alpha}_\beta {(\gam_b)^\rho}_\beta +
   {(\gam_b)^\alpha}_\beta {(\gam_a)^\rho}_\beta = 2 \eta_{ab}
  {\delta^\alpha}_\beta
\lbl{4.29a}
\ee
or shortly,
\be
     \bgam_a \bgam_b + \bgam_b \bgam_a = 2 \eta_{ab} {\bf 1}
\lbl{4.30}
\ee

In particular, using (\ref{4.9}), we find
\be
    \langle {\xi^{\alpha}}^\ddg \gam_0 \xi_\beta \rangle_S =
    \begin{pmatrix}  1 & 0 & 0 & 0 \\
                 0 & 1 & 0 & 0 \\
                 0 & 0 & -1 & 0 \\
                 0 & 0 & 0 & -1 \end{pmatrix} = 
                 {(\gam_0)^\alpha}_\beta = z_{\alpha \beta}
\lbl{4.31}
\ee
The Dirac matrix $\bgam_0 = {(\gam_0)^\alpha}_\beta$ has in this
representation the same form as the spinor metric $z_{\alpha \beta}$.

The quadratic form (\ref{4.21}) explicitly reads
\be
    \langle \Psi^\ddg \Psi \rangle_S = \psi^{*{\tl A}} Z_{{\tl A}{\tl B}}
    \psi^{\tl B} = \psi^{* \alpha i}
     Z_{\alpha i \, \beta j} \psi^{\beta j}
\lbl{4.32}
\ee
The sector of the above expression which belongs to one particular left
ideal is obtained by fixing the ideal indices $i=j$. If, for simplicity,
we omit the latter indices, we obtain for one particular left ideal the
following quadratic form
\be
   \psi^{* \alpha} z_{\alpha \beta} \psi^\beta = \psi_\alpha^* \psi^\alpha
   = {\bar \psi}^\alpha \psi^\alpha
\lbl{4.33}
\ee
where
\be
     {\bar \psi}^\alpha = \psi_\alpha^* = \psi^{* \alpha} z_{\alpha \beta}
\lbl{4.34}
\ee
Since the components of the matrix $z_{\alpha \beta}$ are equal to
the components
of the matric $\bgam_0$, the latter expression corresponds to
the definition of the Dirac adjoint:
\be
    {\bar \psi} = \psi^\dg \bgam_0 \ee
where $\dg$ denotes Hermitian conjugation of a column spinor $\psi$.

We have thus shown the place of the ordinary Dirac (column) spinors and
their adjoints within the framework of such generalized geometric
(Clifford algebra based) approach. The Dirac matrix $\bgam_0$ entering
the definition of the Dirac adjoint spinor corresponds
---in this particular representation--- to the {\it metric}
in the spinor space.

\paragraph{Correspondence with the Dirac bra and ket notation}

The Dirac bra and ket notation is commonly used in quantum theory.
The following relations hold:
\bear
   &\xi^\alpha = |\xi^\alpha \rangle \; , \quad \xi_\alpha = 
   |\xi_\alpha \rangle&
\lbl{4.36a}
\\
   &{\xi^\alpha}^\ddg = \langle \xi^\alpha | \; , \quad 
   {\xi_\alpha}^\ddg = \langle \xi_\alpha |&
\lbl{4.36b} \\
&\langle {\xi^\alpha}^\ddg \xi_\beta \rangle_S = \langle \xi^\alpha|\xi_\beta
\rangle = {\delta^\alpha}_\beta&
\lbl{4.36c}
\\
&\langle {\xi_\alpha}^\ddg \xi_\beta \rangle_S = \langle \xi_\alpha|\xi_\beta
\rangle = z_{\alpha \beta}&
\lbl{4.36d} \\
&\langle \xi^\alpha | = z^{\alpha \beta} \langle \xi_\beta |
= \langle {\bar \xi}_\alpha | = \langle \xi^\alpha \gam_0 |&
\lbl{4.36e}
\\
    &\xi_\alpha {\xi^\alpha}^\ddg = |\xi_\alpha \rangle \langle 
    \xi^\alpha | = {\bf 1}&
\lbl{4.36f}
\\
&\langle {\xi_\alpha}^\ddg A \xi_\beta \rangle_S = 
\langle \xi_\alpha| A |\xi_\beta \rangle = (A)_{\alpha \beta}&
\lbl{4.36g}
\\
&\langle {\xi^\alpha}^\ddg A \xi_\beta \rangle_S = 
\langle \xi^\alpha| A |\xi_\beta \rangle = {(A)^\alpha}_\beta&
\lbl{4.36h}\\
& \Psi = \psi^\alpha \xi_\alpha = |\xi_\alpha \rangle \langle \xi^\alpha|\Psi
      \rangle&
\lbl{4.36i}
\ear
where $\psi^\alpha = \langle {\xi^\alpha}^\ddg \Psi \rangle_S =
\langle \xi^\alpha | \Psi \rangle$.

The above relations hold for one ideal. They can be extended to all four
ideals by replacing the spinor indices $\alpha, \beta$ with the generalized
spinor indices ${\tl A},{\tl B}$.

\subsection{Extending the Dirac equation to curved Clifford space}

In refs.\,\ci{PavsicSchladming,PavsicCliff,PavsicBook} it was
proposed that the
polyvector valued wave function satisfies the Dirac equation in $C$-space:
\be
     \p \Psi \equiv \gam^M \p_M \Psi = 0
\lbl{4.37}
\ee
The latter equation is just the square root of the ``massless" Klein-Gordon
equation in $C$-space, $\p \p \Psi = 0$, considered in ref.\,\ci{PavsicBook},
the scalar part of which was considered by Castro \ci{CastroFound}.
Castro's equation in turn generalizes Pezzaglia's equation \ci{Pezzaglia}
which is based
on the Dixon  \ci{DixonMass} generalization of the Einstein relation for
spinning bodies: $p_\mu p^\mu - S^{\mu \nu} S_{\mu \nu} = m^2$.

The derivative $\p_M$ is the same derivative introduced in
eqs. (\ref{2.15})--(\ref{2.18}). Now it acts on the object
$\Psi$ which, according to eq.\,(\ref{4.10}), is expanded in terms of
the 16 basis elements
$\xi_{\tilde A}$  which, in turn,
can be written as a superposition
of basis elements $\gam_A$ of Clifford algebra. The action of
$\p_M$ on $\gam_A$ is given in (\ref{2.17}). An analogous expression
holds if $\p_M$ operates on
the spinor basis elements $\xi_{\tilde A}$:
\be
    \p_M \xi_{\tilde A} = {{\Gam_M}^{\tilde B}}_{\tl A} \xi_{\tl B}
\lbl{4.38}
\ee
where ${{\Gam_M}^{\tilde B}}_{\tl A}$ are components of the generalized
{\it spin connection},
i.e., the components of the connection of curved
$C$-space for the generalized spinor frame field $\{\xi_{\tl A}\}$.
Using the expansion (\ref{4.10}) and eq.\,(\ref{4.38}) we find
\be
   \p \Psi = \gam^M \p_M (\psi^{\tl A} \xi_{\tl A}) = 
   \gam^M (\p_M \psi^{\tl A}
  + {{\Gam_M}^{\tl A}}_{\tl B} \psi^{\tl B}) \xi_{\tl A} \equiv
    \gam^M \xi_{\tl A} \DD_M \psi^{\tl A} = 0
\lbl{4.39}
\ee
This is just a generalization of the ordinary Dirac equation in
curved spacetime. Instead of curved spacetime, spin connection and
the Dirac spinor, we have now curved Clifford space, generalized
spin connection and the generalized spinor $\psi^{\tl A}$ which 
incorporates 4 independent Dirac spinors, as indicated in eq. (\ref{4.10}).

We may now use the relations (\ref{4.17}),(\ref{4.18}),(\ref{4.24})
and project eq.\,(\ref{4.39}) onto its component form
\be
     {{(\gam^M)}^{\tl C}}_{\tl A} (\p_M \psi^{\tl A} + 
  {{\Gam_M}^{\tl A}}_{\tl B} \psi^{\tl B} ) = 0
\lbl{4.40}
\ee
The spinor indices ${\tl A},~{\tl B}$ can be omitted and eq.
(\ref{4.40}) written simply as
\be
    \bgam^M (\p_M  + {\bf \Gam_M}) \psi = 0
\lbl{4.41}
\ee
where
\be
    \bgam^M = {(\gam^M)^{\tl A}}_{\tl B} \; , \quad 
    {\bf \Gam}_M = {{\Gam_M}^{\tl A}}_{\tl B}
\ee
{\it We see that in the geometric form of the Dirac equation (\ref{4.37})
spin connection is automatically present through the operation
of the derivative $\p_M$ on a polyvector field $\Psi$ written as a
superposition of basis spinors $\xi_{\tl A}$.} The reader has to
be careful (i) not to confuse our symbol $\p_M$, when acting on a
polyvector-valued (Clifford-valued) object,
with a partial derivative, acting on the scalar-valued matrix components
of a polyvector
(see Sec 4.3), (ii) not to miss the fact that $\Psi$ in
eq. (\ref{4.37})
{\it is a Clifford algebra valued object, not just a component
spinor}, and (iii) not hastily think that eq. (\ref{4.37}) lacks
covariance. In fact the simple eq.\,(\ref{4.37}) is equivalent to eq.\,
(\ref{4.39}).

\subsection{The difference between $\p_M \gam_A$ and $\p_M \bgam_A$}

We have already stressed that the action of the derivative $\p_M$ is
different when operating on a scalar or on a Clifford algebra valued object.
We have considered the objects such as $\gam_A$ which form the basis of
Clifford algebra, or $\xi_{\tl A}$ which form the generalized
spinor basis, spanning all four left ideals. The derivatives 
$\p_M \gam_A$ and $\p_M \xi_{\tl A}$ are given in eqs.\,(\ref{2.17}) and
(\ref{4.38}), respectively. In those cases we have applied the derivative
$\p_M$ on the {\it Clifford algebra valued objects} 
$\gam_A$ and $\xi_{\tl A}$.
From the latter objects we can obtain the {\it scalar valued} matrix elements
by using eq.\,(\ref{4.18}). For instance,
\be
    \langle {\xi^{\tl A}}^\ddg \gam_A \xi_{\tl B} \rangle_S =
    {(\gam_A)^{\tl A}}_{\tl B} \equiv \bgam_A
\lbl{4.43}
\ee
\be
    \langle {\xi^{\tl A}}^\ddg \gam_M \xi_{\tl B} \rangle_S =
    {(\gam_M)^{\tl A}}_{\tl B} \equiv \bgam_M
\lbl{4.44}
\ee
Taking the derivative, we have
\bear
     \p_M \bgam_A &=& \p_M \langle {\xi^{\tl C}}^\ddg \gam_A \xi_{\tl D}
     \rangle_S = \langle \p_M {\xi^{\tl C}}^\ddg \gam_A \xi_{\tl D} +
     {\xi^{\tl C}}^\ddg \gam_A \p_M \xi_{\tl D} +
     {\xi^{\tl C}}^\ddg \p_M \gam_A \xi_{\tl D} \rangle_S \nonumber \\
     &=& \langle - {{\Gam_M}^{\tl C}}_{\tl E} \xi^{\tl E}
      \gam_A \xi_{\tl D} +
     {\xi^{\tl C}}^\ddg \gam_A \xi_{\tl E} {{\Gam_M}^{\tl A}}_{\tl D}
     - {{\Omega_A}^B}_M {\xi^{\tl C}}^\ddg \gam_B \xi_{\tl D} \rangle_S
     \nonumber \\
     &=& - {\bf \Gam}_M \bgam_A + \bgam_A {\bf \Gam}_M -
     {{\Omega_A}^B}_M \bgam_B
\lbl{4.45}
\ear
\bear
     \p_M \bgam_N &=& \p_M \langle {\xi^{\tl C}}^\ddg \gam_N \xi_{\tl D}
     \rangle_S = \langle \p_M {\xi^{\tl C}}^\ddg \gam_N \xi_{\tl D} +
     {\xi^{\tl C}}^\ddg \gam_N \p_M \xi_{\tl D} +
     {\xi^{\tl C}}^\ddg \p_M \gam_N \xi_{\tl D} \rangle_S \nonumber \\
     &=& \langle - {{\Gam_M}^{\tl C}}_{\tl E} \xi^{\tl E}
     \gam_N \xi_{\tl D} +
     {\xi^{\tl C}}^\ddg \gam_N \xi_{\tl E} {{\Gam_M}^{\tl A}}_{\tl D} 
     + \Gam_{MN}^J {\xi^{\tl C}}^\ddg \gam_J \xi_{\tl D} \rangle_S
     \nonumber \\
     &=& - {\bf \Gam}_M \bgam_N + \bgam_A {\bf \Gam}_M +
     \Gam_{MN}^J \bgam_J
\lbl{4.46}
\ear
In the above calculation we have distinguished between the derivative of
$\xi_{\tl A}$ and $\xi^{\tl A} = Z^{{\tl A}{\tl B}} \xi_{\tl B}$:
\be
    \p_M \xi_{\tl A} = {{\Gam_M}^{\tl C}}_{\tl A}\, \xi_{\tl C} \; , \quad
    \p_M \xi^{\tl A} = - {{\Gam_M}^{\tl A}}_{\tl C} \, \xi^{\tl C}
\lbl{4.47}
\ee
The latter relations are consistent with
    $$\langle {\xi^{\tl A}}^\ddg \xi_{\tl B} \rangle_S = 
    {\delta^{\tl A}}_{\tl B}$$
\be
  \langle \p_M {\xi^{\tl A}}^\ddg \xi_{\tl B} + 
       {\xi^{\tl A}}^\ddg \p_M \xi_{\tl B} \rangle_S =
       \langle - {{\Gam_M}^{\tl A}}_{\tl C} {\xi^{\tl C}}^\ddg \xi_{\tl B}
       + {\xi^{\tl A}}^\ddg {{\Gam_M}^{\tl C}}_{\tl B} \, \xi_{\tl C}
       \rangle_S = - {{\Gam_M}^{\tl A}}_{\tl B} + 
       {{\Gam_M}^{\tl A}}_{\tl B}
       = 0
\lbl{4.48}
\ee
If we take into account the relation $\gam_N = {e_N}^A \gam_A$ (see
eq.\,(\ref{2.11})) and insert eq.\,(\ref{4.45}) into eq.\,(\ref{4.46}),
then we obtain
\be
     (\p_M {e_N}^A - \Gam_{MN}^J {e_J}^A - {{\Omega_B}^A}_M {e_N}^B)
     \bgam_A = 0
\lbl{4.49}
\ee
From the latter equation it follows
\be
    \p_M {e_N}^A - \Gam_{MN}^J {e_J}^A - {{\Omega_B}^A}_M {e_N}^B = 0
\lbl{4.50}
\ee
In eqs.\,(\ref{4.46}),(\ref{4.50}) we recognize a generalization of
the corresponding relations for the Dirac matrices and vielbein in
curved spacetime, namely
\be
    \p_\mu \bgam_\nu - \Gam_{\mu \nu}^\rho \bgam_\rho =
    [\bgam_\nu, {\bf \Gam}_\mu ]
\lbl{4.51}
\ee  \be
   \p_\mu {e_\nu}^a - \Gam_{\mu \nu}^\rho  {e_\rho}^a
    - {{\omega_b}^a}_\mu {e_\nu}^b = 0
\lbl{4.52}
\ee
In fact, the latter relations are just a particular case of the relations
(\ref{4.46}),(\ref{4.50}). If we restrict our general geometric spinors
to one ideal only, and assume that the vielbein and the affinity of
$C$-space do not mix all 16 components of the Clifford algebra, but
only those four that belong to spacetime, we obtain just the ordinary
theory of spinors in curved spacetime.

The procedure that we have developed here demonstrates that the symbol
$\p_M$ for the derivative indeed cannot be confused with some other derivative
operator, because there are no other derivative operators within the
framework of our formalism\footnote{In ref.\,\ci{PavsicOrder} we
distinguished between two sorts of derivative, but in Sec.\,4.3 we
have seen that  there are in fact only two sorts of objects, namely
$\gam_M$ and matrices $\bgam_M$.}.
We have shown that when the derivative operates on the matrices, e.g.,
the Dirac matrices, we obtain the expressions (\ref{4.46})--(\ref{4.52})
which are different from the expressions (\ref{2.16}),(\ref{2.17}),
(\ref{4.47}) which hold when $\p_M$ operates on the Clifford algebra
valued objects $\gam_M,~\gam_A,~\xi_{\tl A}$, etc. In short, there is
a clear distinction between the abstract Clifford numbers and their
representation by matrices. This becomes manifest, e.g., under the action
of the derivative $\p_M$.

\section{Yang-Mills gauge fields as the spin connection in $C$-space}

\subsection{Local (pseudo) rotations in $C$-space}

Let us define the generators of local (pseudo) rotations
in a tangent $C$-space $T_X (C)$ according to
\be
      \Sigma_{AB} = - \Sigma_{BA}
      = \left\{  \begin{array}{ll}
      ~~\gam_A \gam_B\; , & {\rm if} ~A < B \\
      - \gam_A \gam_B\; , & {\rm if} ~A > B \\
      0 \; , & {\rm if} ~A = B
      \end{array}
      \right.
\lbl{5.1}
\ee
Here we assume that the basis elements $\gam_A,~\gam_B$ and the indices
$A,~B$ are ordered according to
the rule suggested in eq.\,(\ref{2.2}).

We also have $\Sigma_{AB} = {f_{AB}}^C \gam_C$, where ${f_{AB}}^C$
are constants. Remember that by ${\bf o}$ we denote a scalar component of
a polyvector, so that for
$C= {\bf o}$ we have $\gam_{\bf o} \equiv {\bf 1}$.

A generic transformation in a tangent $C$-space $T_X (C)$ which maps
a polyvector
 $\Psi$ into another polyvector $\Psi'$ is given by
\be
   \Psi' = R \Psi S
\lbl{5.2}
\ee
where 
\be
     R = {\rm e}^{{1\oo 4} \Sigma_{AB} \alpha^{AB}}
     = {\rm e}^{\gam_A \alpha^A}
     \quad {\rm and}
     \quad S = {\rm e}^{{1\oo 4} \Sigma_{AB} \beta^{AB}} 
     = {\rm e}^{\gam_A \beta^A}
\lbl{5.3}
\ee
Here $\alpha^{AB}$ and $\beta^{AB}$, or equivalently, 
$\alpha^A = {f_{CD}}^A \alpha^{CD}$ and $\beta^A = {f_{CD}}^A \beta^{CD}$,
are parameters of the transformation.

In general, eq.\,(\ref{5.2}) allows for the transformation which maps a
basis element $\gam_A$ into a mixture of basis elements:
\be
    \gam_A \rightarrow \gam'_A = R \gam_A S = {L_A}^B \gam_B
\lbl{5.3a}
\ee

In particular, we
have the following three interesting cases:

      (i) $\alpha^{AB} \neq 0, ~~\beta^{AB} = - \alpha^{AB}$. Then we have
\be
    \Psi' = R \Psi R^{-1}
\lbl{5.4}
\ee
This is the transformation which preserves the Clifford algebra relations
\be
     [\gam_A,\gam_B] = {C_{AB}}^C \gam_C
\lbl{5.5}
\ee
so that for the transformed elements $\gam'_A = R \gam_A R^{-1}$ we have
\be
     [\gam'_A,\gam'_B] = {C_{AB}}^C \gam'_C
\lbl{5.6}
\ee
with the same structure constants ${C_{AB}}^C$. This means that the
transformation (\ref{5.4}) maps one basis element $\gam_A$ into another
basis element $\gam'_{A}$, e.g., a basis vector into a bivector,
a 3-vector into 1-vector, etc.

    (ii) $\alpha^{AB} \neq 0, ~~\beta^{AB} = 0$. Then we have
\be
     \Psi' = R \Psi
\lbl{5.7}
\ee
This is the transformation which maps a basis spinor $\xi_{\alpha i}$ into
another basis spinor $\xi'_{\alpha i}$ belonging to the same left ideal:
\be
    \xi_{\alpha i} \subset {\cal I}_i^L  \rightarrow \xi'_{\alpha i} =
    R \xi_{\alpha i}
   \subset {\cal I}_i^L
\lbl{5.8}
\ee

   (iii) $\alpha^{AB} = 0, ~~\beta^{AB} \neq 0$. Then
\be
    \Psi' = \Psi S
\lbl{5.9}
\ee
This is the transformation that maps a right ideal into the same right ideal:
\be
    \xi_{\alpha i} \subset {\cal I}_i^R  \rightarrow \xi'_{\alpha i} =  
    \xi_{\alpha i} S
   \subset {\cal I}_i^R
\lbl{5.10}
\ee

In general, for the transformation (\ref{5.2}) we have
\be
    \Psi' = \psi^{\tl A} R \xi_{\tl A} S = 
    \psi^{\tl A} {U_{\tl A}}^{\tl B} 
    \xi_{\tl B}
  = \psi'^{\tl A} \xi_{\tl A}
\lbl{5.11}
\ee
where
\be
    \psi'^{\tl A} = {U^{\tl A}}_{\tl B} \psi^{\tl B}
\lbl{5.12}
\ee
The latter transformations, in general, mixes right and left ideals.
Eq.\,(\ref{5.12})
can be considered as a matrix equation in the space spanned by the generalized
spinor indices ${\tl A},~{\tl B}$:
\be
       \psi' = {\bf U} \psi
\lbl{5.13}
\ee
where ${\bf U}$ is a $16 \times 16$ matrix, whilst $\psi$ and $\psi'$
are columns with 16 elements. 

It is illustrative to calculate the matrix elements of $\Psi'$ with respect
to the spinor basis of, say, the first left ideal $\xi^\alpha \equiv
\xi^{\alpha 1}$. So we obtain matrices $\psi'^{\alpha \beta}$ and
$\psi^{\alpha \beta}$ representing the polyvectors $\Psi'$ and $\Psi$
respectively;
the index $\beta$ says which column, i.e., which left ideal
(it stands now instead of the index $i$ or $j$):
\bear
    \langle {\xi^\gam}^\ddg \Psi' \xi^\delta \rangle_S &=&
    \langle {\xi^\gam}^\ddg R \Psi S \xi^\delta \rangle_S =
    \langle {\xi^\gam}^\ddg R \xi_\alpha {\xi^\alpha}^\ddg 
    \Psi \xi^\beta \xi_\beta^\ddg   S \xi^\delta \rangle_S \nonumber \\
    &=& {R^\gam}_\alpha \, \psi^{\alpha \beta} \,{S_\beta}^\delta =
    {U^{(\gam \delta)}}_{(\alpha \beta)} \, \psi^{(\alpha \beta)} =
    {U^{\tl B}}_{\tl C} \, \psi^{\tl C}
\lbl{5.13a1}
\ear
where
\be 
   {U^{\tl B}}_{\tl C} \equiv {U^{(\gam \delta)}}_{(\alpha \beta)} =
   {R^\gam}_\alpha {S_\beta}^\delta
\lbl{5.13a2}
\ee
In matrix notation this reads
\be
{\bf U} = {\bf R} \otimes {\bf S}^{\rm T}
\lbl{5.14}
\ee
where ${\bf R}$ and ${\bf S}$ are
$4 \times 4$ matrices representing the Clifford numbers $R$ and $S$.
That is, ${\bf U}$ is the direct product of ${\bf R}$ and the transpose 
${\bf S}^{\rm T}$
of ${\bf S}$, and it belongs, in general, to the group $GL(4,C) \times GL(4,C)$.
The group is local, because the basis elements $\gam_A$ entering the
definition (\ref{4.14}) depend on position $X$ in $C$-space according
to the relations (\ref{2.17}), and also the group parameters
$\alpha^A,~\beta^A$ in general depend on $X$.  

The most general gauge group here is GL(4,C) $\times$ GL(4,C). The first
piece belongs to the left transformations $R$ and the second piece to the
right transformations $S$ of eq.\,(\ref{5.2}).
The group GL(4,C)$\times$ GL(4,C) that we started from is subjected
to further restrictions resulting from the requirement that the transformations
(\ref{5.2}) should leave the quadratic form $\Psi^{\ddg} * \Psi$ invariant.
So we have
\be
   \Psi'^\ddg * \Psi' = \langle \Psi'^\ddg \Psi' \rangle_S =
\langle S^\ddg \Psi^\ddg R^\ddg R \Psi S \rangle_S = 
\langle \Psi^\ddg \Psi \rangle_S = \Psi^\ddg * \Psi
\lbl{5.26}
\ee
provided that
\be
    R^\ddg R = {\bf 1} \quad  {\rm and} \quad S^\ddg S = {\bf 1}.
\lbl{5.27}
\ee

Using the exponential expression (\ref{5.3}) we have
\be
    R = {\rm e}^{\gam_A \alpha^A} \; , \quad R^{-1} = {\rm e}^{-\gam_A \alpha^A}
\lbl{5.27a}
\ee
\be
    R^\ddg = \left ({\rm e}^{\gam_A \alpha^A} \right )^\ddg =
    (1 + \gam_A \alpha^A + {1\oo 2!} (\gam_A \alpha^A)^2 + ...)^\ddg
    {\rm e}^{(\gam_A \alpha^A)^\ddg} = {\rm e}^{\gam_A^\ddg \alpha^{*A}}
\lbl{5.27b}
\ee
where, according to our definition, reversion acts also on imaginary
number $i$.

After taking into account eqs.\,(\ref{5.27a}),(\ref{5.27b}), the condition
(\ref{5.27}) which reads $R^\ddg = R^{-1}$ transforms into
\be
    \gam_A^\ddg \alpha^{*A} = - \gam_A \alpha^A
\lbl{5.27c}
\ee
Since
\be
     \gam_A^\ddg = \gam_A \quad {\rm if} \quad A \in \lbrace A \rbrace_+
     \; , \quad \gam_A^\ddg = 
     - \gam_A \quad {\rm if}\quad {\rm if} \quad A \in  
     \lbrace A \rbrace_-
\lbl{5.27d}
\ee
we have that $\gam_A^\ddg$ is equal to plus or minus $\gam_A$, depending
on the grade, i.e., on the type of the multivector index $A$, and consequently
the parameters $\alpha^A$ are either real or imaginary\footnote{
Later we will show (see eq.\,(\ref{5.49a2})) that it is convenient to redefine
the Clifford basis so that its elements are invariant under reversion:
$\gam_A^\ddg = \gam_A$. Then the parameters $\alpha^A$ can be kept
{\it imaginary} for all $A$, if the form
$R={\rm exp} \, [\gam_A \alpha^A]$ is
used, or {\it real}, if the form 
$R={\rm exp} \, [i \, \gam_A \alpha^A]$ is used.}.

If we represent the Clifford numbers $R,~S$ by $4 \times 4$ matrices
\be
    {\bf R} = {R^\alpha}_\beta = \langle {\xi^\alpha}^\ddg R \xi_\beta
    \rangle_S \; , \quad 
    {\bf S} = {S^\alpha}_\beta = \langle {\xi^\alpha}^\ddg S \xi_\beta
    \rangle_S
\lbl{5.28}
\ee
where $\xi^\alpha\equiv \xi^{\alpha 1}, ~\xi^{\beta} \equiv \xi^{\beta 1}$
then the condition (\ref{5.27}) reads
\be
   \langle {\xi^\alpha}^\ddg R^\ddg R \xi_\beta \rangle_S
   ={\delta^\alpha}_\beta \quad {\rm and} \quad
   \langle {\xi^\alpha}^\ddg S^\ddg S \xi_\beta \rangle_S
   ={\delta^\alpha}_\beta
\lbl{5.29}
\ee
which, after lowering the index $\alpha$ with the metric $z_{\alpha \beta}$,
becomes
\be
    \langle {\xi_\alpha}^\ddg R^\ddg R \xi_\beta \rangle_S
   = z_{\alpha \beta}
\lbl{5.30}
\ee
Using $\xi_\delta {\xi^\delta}^\ddg =\xi^\delta {\xi_\delta}^\ddg = 
\xi^\delta z_{\delta\gam} {\xi^\gam}^\ddg  ={\bf 1}$
we have
\be
    \langle {\xi_\alpha}^\ddg R^\ddg 
    \xi^\delta z_{\delta\gam}{\xi^\gam}^\ddg R \xi_\beta \rangle_S
    = \langle {\xi_\alpha}^\ddg R^\ddg \xi^\delta \rangle_S \, z_{\delta\gam}\,
    \langle {\xi^\gam}^\ddg R \xi_\beta \rangle_S = z_{\alpha\beta}
\lbl{5.31}
\ee
This can be written as
\be
   \langle ({\xi^\delta}^\ddg R \xi_\alpha)^\ddg \rangle_S \, z_{\delta\gam}
    \langle {\xi^{\gamma}}^\ddg R \xi_\beta \rangle_S = z_{\alpha\beta}
\lbl{5.32}
\ee
or
  \be    {\bf R}^\dg {\bf z} {\bf R} = {\bf z}
\lbl{5.33}
\ee
where ${\bf R}^\dg = \langle ({\xi^\delta}^\ddg R \xi_\alpha)^\ddg \rangle_S$
is the Hermitian conjugate of matrix ${\bf R} =
 \langle {\xi^\delta}^\ddg R \xi_\alpha \rangle_S$
 
The transformations (\ref{5.2}), in particular, transform a scalar,
or a vector
field into a spinor field. In this respect they have the role of
supersymmetric transformations. Supersymmetry is automatically
present in our approach. A detailed investigation of this fascinating
possibility is beyond the scope of this paper. A different approach
to supersymmetry and Clifford algebra was proposed in refs.\,\ci{CastroSuper}.

While reversion $\ddg$ refers to a Clifford valued object, say $A$,
hermitian conjugation $\dg$ refers to the {\it matrix} representing $A$.
So we have
\be
    \langle {\xi^\alpha}^\ddg A \xi_\beta \rangle_S = {A^\alpha}_\beta
    = {\bf A}
\lbl{5.33a}
\ee
\be
    \langle ({\xi^\alpha}^\ddg A \xi_\beta)^\ddg \rangle_S =
    \langle \xi_\beta^\ddg A^\ddg \xi^\alpha \rangle_S = {A_\beta^*}^\alpha
    = {\bf A}^\dg    
\lbl{5.33b}
\ee
Complex conjugation $*$ comes from the reversion which, according
to our definition operates also on the (commuting) imaginary unit $i$
(the bivector of phase space) entering the definition of $\xi_\alpha$
(see eq.\,(\ref{4.9})).

\subsection{Invariance of the generalized Dirac equation under local
(pseudo) rotations in $C$-space}

We will now explore the invariance properties of the generalized Dirac equation
(\ref{4.37}). An $X$-dependent polyvector valued field can be expanded
either in terms of the Clifford algebra basis $\lbrace \gam_A \rbrace$,
or in
terms of the ideal basis $\lbrace \xi_{\tl A} \rbrace$, or in terms of the
$C$-space coordinate basis $\lbrace \gam_M \rbrace$,
defined
by eq.\,(\ref{2.11}). For illustration, let us first consider
the latter case. Using (\ref{2.16}),(\ref{2.18}), we have
\be
    \p \Phi = \gam^M \p_M (\phi^N \gam_N) = \gam^M (\p_M \phi^N +
    \Gam_{MJ}^N \phi^J) \gam_N = \gam^M \gam^N \, \DD_M \phi^N
\lbl{5.15}
\ee
where $\DD_M$ is the covariant derivative.

In another coordinate basis $\lbrace \gam'_M \rbrace$ we have
\be
    \p' \Phi' = \gam'^M \p'_M (\phi'^N \gam'_N) = 
    \gam'^M (\p'_M \phi'^N +
    {\Gam'}_{MJ}^N \phi'^J) \gam'_N = \gam'^M \gam'^N \, \DD'_M \phi'^N
\lbl{5.16}
\ee

The new and old basis elements and components are related by general
coordinate transformations in $C$-space:
\be
    \phi'^N = {C^N}_J \phi^J \; , \quad \gam'_N = {C^J}_N \gam_J
\lbl{5.17}
\ee
where ${C^N}_J \equiv \p x'^N/\p x^J$. Requiring 
\be
    \p \Phi = \p' \Phi'
\lbl{5.18}
\ee
and using eqs.\,(\ref{5.15}),(\ref{5.16}) we find the following
transformation law for the $C$-space affine connection:
\be
    \Gam_{MN}^J = {C_M}^R {C_N}^S {C^J}_K {\Gam'}_{RS}^K + \p_M
    {C^R}_N \, {C_R}^J
\lbl{5.19}
\ee
The latter relation generalizes the well known transformation property
of the affine connection.

Similarly, for $\Phi = \Psi = \psi^{\tl A} \xi_{\tl A}$ we have
\be
    \p' \Psi' = \p \Psi = \gam'^M \p'_M (\psi'^{\tl A} \xi'_{\tl A}) =
    \gam^M \p_M (\psi^{\tl A} \xi_{\tl A})
\lbl{5.20}
\ee
Using eqs.\,(\ref{4.38}) and (\ref{5.12}) and
\be
     \p'_M \xi'_{\tl A} = 
     {{\Gam'_M}^{\tl B}}_{\tl A} \, \xi'_{\tl B} \quad 
     {\rm and} \quad
     \xi'_{\tl A} = {U^{\tl C}}_{\tl A} \xi'_{\tl C}
\lbl{5.22}
\ee
we obtain
\be
     {\Gamma_{M {\tl A}}}^{\tl B} = {C^M}_N {U_{\tl D}}^{\tl B}
     {U^{\tl C}}_{\tl A} {\Gamma'_{M {\tl C}}}^{\tl D} +
      \p_M {U^{\tl D}}_{\tl A} \, {U_{\tl D}}^{\tl B}
\lbl{5.21}
\ee
which is the transformation law for the generalized spin connection
(i.e., the spin
connection in $C$-space). The relation (\ref{5.21}) is analogous
to the relation (\ref{5.19}).

In a special case when we do not perform a general coordinate transformation
(\ref{5.17}), but only a local (pseudo) rotations (\ref{5.12}),(\ref{5.22}),
we have
\be
    {\Gamma_{M {\tl A}}}^{\tl B} = {U_{\tl D}}^{\tl B}
     {U^{\tl C}}_{\tl A} {\Gamma'_{M {\tl C}}}^{\tl D} +
      \p_M {U^{\tl D}}_{\tl A}\, {U_{\tl D}}^{\tl B}
\lbl{5.23}
\ee
In  matrix notation\footnote{The objects
are considered as matrices in the generalized spinor indices ${\tl A},~
{\tl B},~{\tl C},~{\tl D}$.} this reads
\be
    {\bf \Gamma}_M = {\bf U} {\bf \Gamma'_M} {\bf U}^{-1} + 
    {\bf U} \p_M {\bf U}^{-1}
\lbl{5.24}
\ee
By renaming, respectively, ${\bf \Gam_M}$, ${\bf U}$ into
${\bf \Gam'_M}$, ${\bf U}^{-1}$, and
vice versa, eq.\,(\ref{5.24}) assumes a more familiar form
\be
    {\bf \Gamma'}_M = {\bf U}^{-1} {\bf \Gamma_M} {\bf U} + 
    {\bf U}^{-1} \p_M {\bf U}
\lbl{5.25}
\ee

We see that ${\bf \Gam}_M$ transforms as a non abelian gauge field.
We have thus demonstrated that the generally covariant Dirac equation in
16-dimensional curved $C$-space contains the coupling of spinor
fields $\psi^{\tl A}$ with non abelian gauge fields
${{\Gam_M}^{\tl A}}_{\tl B}$ which
altogether form the spin connection in $C$-space.
The theory presented here is essentially a generalization of the one considered
by Pezzaglia \ci{Pezzaglia}, where the higher multivector derivatives
were not included, and the geometric interpretation of the imaginary
unit $i$ was different from ours. Pezzaglia in turn takes Crawford's ideas
\ci{Crawford}, but interprets them geometrically.

In eqs.\, (\ref{5.18}),(\ref{5.20}) we considered {\it passive transformations},
that is the transformations which change components {\it and} the basis
elements, so that
\bear
  &\Phi' = \phi'^M \gam'_M = \phi^M \gam_M = \Phi&
\lbl{5.34a}
\\
   &\Psi' = \psi'^{\tl A} \xi'_{\tl A} = \psi^{\tl A} \xi_{\tl A} = \Psi&
\lbl{5.34b}
\ear
Similarly for the derivative:
\be
    \p' = \gam'^M \p'_M = \gam^M \p_M
\lbl{5.34c}
\ee
The invariance of the Dirac equation, written in the geometric form
$\p \Psi = 0$, is straightforward, since the polyvector objects $\p$ and
$\Psi$, which have values in Clifford algebra, do not change under
passive transformations.

Under {\it active transformations} either the components or the basis elements
change:
\be
    \Psi' = \psi^{\tl A} \xi'_{\tl A}
     = \psi^{\tl A} {U^{\tl C}}_{\tl A} \xi_{\tl C}
    = \psi'^{\tl C} \xi_{\tl C}
\lbl{5.35}
\ee
where
\be
   \xi'_{\tl A} = {U^{\tl C}}_{\tl A} \xi_{\tl C}
\lbl{5.36a}
\ee
and
\be
    \psi'^{\tl C} = {U^{\tl C}}_{\tl A} \psi^{\tl A}
\lbl{5.36b}
\ee
Under the transformation (\ref{5.36b}) we have
\be
      {\DD}'_{M} \psi'^{\tl A} =
      {{\p x^N}\oo {\p x'^M}} \, {U^{\tl A}}_{\tl B}
      \, \DD_N \psi^{\tl B}
\lbl{5.37}
\ee
where
\be
      \DD'_{M}\psi'^{\tl A} = \p'_M \psi'^{\tl A} +
      {{\Gam'_{M}}^{\tl A}}_{\tl B} \psi'^{\tl B} \quad {\rm and}\quad
      \DD_{M}\psi^{\tl A} = \p_M \psi^{\tl A} +
      {{\Gam_{M}}^{\tl A}}_{\tl B} \psi^{\tl B}
\lbl{5.38}
\ee
are. respectively, the transformed and the ``original" covariant derivative,
defined in eq.\,(\ref{4.39}). For the reason of completeness, we have
considered in eq.\,(\ref{5.37}) also the general coordinate transformations
of the $C$-space coordinates $x^M=(s,x^\mu,x^{\mu \nu},...)$. The latter
transformations are independent from local transformations, such as
(\ref{5.36a}) or (\ref{5.36b}) which affect the generalized spinor indices
${\tl A},~{\tl B},...$, or from an analogous local transformation which
affects the local (flat) polyvector indices $A,~B,...$\,. If we perform
only a local transformation (\ref{5.36b}) and no general coordinate
transformation, i.e., if we take $\p x^N/\p x'^M = {\delta^N}_M$, then
the transformation of the covariant derivative (\ref{5.37}) reads simply
\be
 {\DD}'_{M} \psi'^{\tl A} = {U^{\tl A}}_{\tl B} \, \DD_M \psi^{\tl B}
\lbl{5.39}
\ee   
The covariant derivative transforms in the same way as the field
$\psi^{\tl A}$.
This is the well known property of gauge theories.

{\it Action}

The $C$-space Dirac equation (\ref{4.37}) can be derived from the action
\be
    I[\Psi, \Psi^\ddg] = \int \dd^{2^n} X \, \sqrt{|G|} \, 
    i \Psi^\ddg \p \Psi = \int \dd^{2^n} X \, \sqrt{|G|} \, i
   {\psi^*}^{\tl B} \xi_{\tl B}^\ddg \gam^M \xi_{\tl A} \DD_M \psi^{\tl A}
\lbl{5.40}
\ee
where $\dd^{2^n} X \, \sqrt{|G|}$ is the invariant volume element of the
$2^n$-dimensional $C$-space, $G\equiv {\rm det} \,\, G_{MN}$ being the
determinant of the $C$-space metric.

Taking the scalar part of the action (\ref{5.40}) we obtain the matrix
form of the action:
\be
    I[{\psi^*}^{\tl B},\psi^{\tl A}] = \int \dd^{2^n} X \, \sqrt{|G|} \,i
    {\psi^*}^{\tl B} (\gam^M)_{\tl B \tl A} \p_M \psi^{\tl A}
\lbl{5.40a}
\ee
where $(\gam^M)_{\tl B \tl A} = 
\langle \xi_{\tl B}^\ddg \gam^M \xi_{\tl A} \rangle_S$.
The quadratic form under the integral can be written in the
form
\be
    {\psi^*}^{\tl B} Z_{\tl B \tl C} {(\gam^M)^{\tl C}}_{\tl A} \, 
    \p_M \psi^{\tl A} = {\psi^*}_{\tl C} {(\gam^M)^{\tl C}}_{\tl A} \, 
    \p_M \psi^{\tl A} = {\bar \psi} \bgam^M \p_M \psi
\lbl{5.40b}
\ee
Here $\psi \equiv \psi^{\tl A}$ represents a column of {\it contravariant}
components with the generalized spinor index ${\tl A}$, while
${\bar \psi} \equiv {\psi^*}_{\tl C}$ represents a row of complex
conjugate {\it covariant} components. Writing the indices ${\tl A},~
{\tl B},...$ in the from ${\tl A}= \alpha i$, ${\tl B} = \beta j$,
and fixing the ideal indices $i,~j$ to one chosen ideal, e.g.,
$i=j=1$ and omitting the index 1, then we find that eq.\,(\ref{5.40b})
contains the ordinary spinor quadratic form
\be
     {\psi^*}_\alpha {(\gam)^\alpha}_{\beta} \p_M \psi^\beta =
     {\bar \psi} \bgam^M \p_M \psi
\lbl{5.40c}
\ee
If $M= \mu,~\mu=0,1,2,3$, we obtain the form 
${\bar \psi} \bgam^\mu \p_\mu \psi$ entering the ordinary Dirac action.

The action (\ref{5.40}) is invariant under {\it passive transformations}
(\ref{5.34b}), and passive general coordinate transformations
$x^M \rightarrow x'^M(x^N)$ (which lead to (\ref{5.34a})--(\ref{5.34c})).
Passive transformations do not transform a geometric object,
such as $\Psi$ or $\p \Psi$. Since they transform the basis
elements and the corresponding components,
for instance $\xi_{\tl A}$ and $\psi^{\tl A}$, they only show how
the components look in different reference frames.

We will now show that the scalar part of the action (\ref{5.40}) is invariant
under the {\it active transformations} as well. We have
\be
   \langle \Psi^\ddg \gam^C {e_C}^M \p_M \Psi \rangle_S \rightarrow
   \langle \Psi'^\ddg \gam'^C {e_C}^M \p_M \Psi' \rangle_S
   = \psi^{* {\tl B}} \langle {\xi'_{\tl B}}^\ddg \gam'^C \xi'_{\tl A}
   \rangle_S \, {e_C}^M \, \DD_M \psi^{\tl A}
\lbl{5.41}
\ee
Examining the matrix elements we find
\bear
   \langle {\xi'_{\tl B}}^\ddg \gam'^C \xi'_{\tl A} \rangle_S &=&
   \langle \xi_{\tl C}^\ddg \, {U^{* {\tl C}}}_{\tl B} \, \gam'^C
   {U^{\tl D}}_{\tl A} \, \xi_{\tl D} \rangle_S =
   {U^{* {\tl C}}}_{\tl B} {(\gam'^C )}_{{\tl C}{\tl D}} \, 
   {U^{\tl D}}_{\tl A}
   ={U^{* {\tl C}}}_{\tl B} {(\gam^A )}_{{\tl C}{\tl D}} 
   {U^{\tl D}}_{\tl A}
   {L_A}^C \nonumber \\
    &=& {(\gam^C )}_{{\tl B}{\tl A}} =
   \langle \xi_{\tl B}^\ddg \gam^C \xi_{\tl A} \rangle_S
\lbl{5.42}
\ear
Here ${L_A}^C$ is a local ``rotation" (i.e., a Lorentz transformation
in $C$-space) acting on local basis elements $\gam^C$. On the other hand
\be
\langle {\xi'_{\tl B}}^\ddg \gam'^C \xi'_{\tl A} \rangle_S
=
\langle {S^\ddg \xi_{\tl B}}^\ddg R^\ddg \gam'^C R\xi_{\tl A} S \rangle_S
= \langle \xi_{\tl B}^\ddg \gam^C \xi_{\tl A} \rangle_S =
{(\gam^C )}_{{\tl B}{\tl A}}
\lbl{5.42aa}
\ee
In the last step of the above equation we have taken into account the
fact that the operation $\langle ~~ \rangle_S$ (to be distinguished from
the operation $\langle ~~ \rangle_0$) leaves the cyclically permutatated
product of geometric
objects (Clifford numbers) invariant under a transformation (\ref{5.2}),
then we used $S S^\ddg=1$ (postulated in Sec.\,5.1), and 
$R^\ddg \gam'^C R = \gam^C$.

Comparing equations
(\ref{5.42}) and (\ref{5.42aa}), after
raising  ${\tl B}$ and lowering ${\tl C}$, we have
\be
  {{U^*}_{\tl C}}^{\tl B} {(\gam^A )^{\tl C}}_{\tl D} {U^{\tl D}}_{\tl A}
   {L_A}^C = {(\gam^C )^{\tl B}}_{\tl A}
\lbl{5.42a}
\ee
i.e.,
\be
   {\bf U}^{-1} \bgam^A {\bf U} \, {L_A}^C = \bgam^C
\lbl{5.42b}
\ee
The latter relation is a generalization of the well known transformation
properties of gamma matrices under Lorentz transformations. 
Writing
 ${\tl A}= \alpha i$, ${\tl B} = \beta j$,
and fixing the ideal indices $i,~j$ to one chosen ideal, e.g.,
$i=j=1$, and omitting index 1, we obtain for $A=a$, $a=0,1,2,3$,
\be
  {{U^*}_\gamma}^{\beta} {(\gam^a )^{\gamma}}_{\delta} 
  {U^{\delta}}_{\alpha}
   {L_a}^c = {(\gam^c )^{\beta}}_{\alpha}
\lbl{5.42c}
\ee    
which is just the ordinary relation.

Using eq.\,(\ref{5.42}) we find that the quadratic form (\ref{5.41}) is
invariant under active transformations.

Alternatively, we may transform the components, while keeping the basis
elements unchanged:
\bear
    \langle \Psi'^\ddg \p' \Psi' \rangle_S &=& \langle \psi'^{*{\tl B}}
    \xi^\ddg_{\tl B} \gam^A \xi_{\tl A} \DD'_A \psi'^{\tl A} \rangle_S
    = {\psi^*}^{\tl D} {{U^*}^{\tl B}}_{\tl D} (\gam^A )_{{\tl B}{\tl A}}\,
    {U^{\tl A}}_{\tl C}\, \DD'_A \psi^{\tl C} \lbl{5.45} \\
     &=&{\psi^*}^{\tl D} {{U^*}^{\tl B}}_{\tl D} (\gam^A )_{{\tl B}{\tl A}}
    {U^{\tl A}}_{\tl C} {L_A}^C \DD_C \psi^{\tl C} =
    \psi^{*{\tl D}} (\gam^C )_{{\tl D}{\tl C}} \DD_C \psi^{\tl C}
    = \langle \Psi^\ddg \p \Psi \rangle_S
\nonumber
\ear
Above we have used
\be
    \gam^M \DD_M \psi^{\tl A} = {e_A}^M \gam^A \DD_M \psi^{\tl A} =
    \gam^A \DD_A \psi^{\tl A} \rightarrow \gam^A \DD'_A \psi^{\tl A}
\lbl{5.46}
\ee
and
 \be  {\DD'}_A \psi'^{\tl A} = {L_A}^C \DD_C \psi'^{\tl A}
\lbl{5.47}
\ee
We also used eq.\,(\ref{5.42a}).

Eqs.\.(\ref{5.41})--(\ref{5.42b}) and (\ref{5.45}) demonstrate the
invariance of the scalar part of the action (\ref{5.40}) under
active local rotations in $C$-space.

{\it Noether's current}

Let us perform the variation of the action (\ref{5.40}) by retaining the
term which results from the variation of the boundary. We obtain:
\bear
    \delta I &=& \int \dd^{2^n} X \, i \left \lbrace \sqrt{|G|}
    (\delta \Psi^\ddg \gam^M \p_M \psi + 
    \Psi^\ddg \gam^M \p_M \delta \Psi +
    \p_M (\sqrt{|G|} \Psi^\ddg \gam^N \p_N \Psi \delta x^M) \right \rbrace
      \nonumber\\
    &=& \int \dd^{2^n} X \, i \left \lbrace \sqrt{|G|}
    \delta \Psi^\ddg \gam^M \p_M \psi - \p_M (\sqrt{|G|} \, \Psi^\ddg \gam^M)
    \delta \Psi \right . \nonumber \\
    &&\hs{2cm} + \left . \p_M \left [ \sqrt{|G|} 
    (\Psi^\ddg \gam^M \delta \Psi +
    \Psi^\ddg \gam^N \p_N \Psi \, \delta x^M ) \right ] \right \rbrace
\lbl{5.48}
\ear

If we fix the boundary, i.e., if we take $\delta x^M = 0$, then from
eq.(\ref{5.48}) we read the equations of motion:
\be
      \gam^M \p_M \Psi = \gam^A {e_A}^M \, \p_M \Psi =   0
\lbl{5.48a}
\ee
and
\be
     (\p_M \Psi^\ddg)\, \gam^M =  {e_A}^M \p_M \Psi^\ddg \gam^A =  0
\lbl{5.48b}
\ee
where we have used $\p_M (\sqrt{|G|} \gam^M ) = 0$.
Now, performing the operation of reversion on equation (\ref{5.48a})
we obbtain
\be
     (\p_M \Psi^\ddg)\, (\gam^M)^\ddg = 
     {e_A}^M \p_M \Psi^\ddg (\gam^A)^\ddg = 0 
\lbl{5.48c}
\ee
which is not the same as eq.\,(\ref{5.48b}). The equations of motion for
$\Psi$ and its reverse, $\Psi^\ddg$, are therefore only consistent
if all  $\gam^M$'s entering the equations satisfy either (i)
$(\gam^M)^\ddg =
\gam^M$, or (ii)  $(\gam^M)^\ddg = -\gam^M$. Expanding 
$\gam^M = {e_A}^M \gam^A$ we see that only those Clifford basis
numbers $\gam^A$ which are either even ($(\gam^A)^\ddg = \gam^A$),
or odd ($(\gam^A)^\ddg = - \gam^A$) under reversion can simultaneously
enter the equations of motion. From our action (\ref{5.40}) we thus
obtain two different classes of equations of motion, one class for
those values
of the multivector indices $A$ for which $\gam^A$ is even, and the
other class
for those $A$ for which $\gam^A$ is odd under reversion. This is so,
because our action is not invariant under reversion.

An alternative is to consider the action which is invariant under reversion:
\be
    I[\Psi, \Psi^\ddg] = {1\oo 2} \int \dd^{2^n} X \, \sqrt{|G|} \, 
    \left [
    i \Psi^\ddg \p \Psi + (i \Psi^\ddg \p \Psi)^\ddg \right ]
\lbl{5.48d}
\ee
Then the corresponding equations of motion are
\be
       (\gam^M + (\gam^M)^\ddg ) \p_M \Psi = 0 \quad {\rm and} \quad
       \p_M \Psi^\ddg (\gam^M + (\gam^M)^\ddg ) = 0
\lbl{5.48e}
\ee
Under reversion the first of equations (\ref{5.48e}) is transformed into
the other, and vice versa. Therefore the equations of motion (\ref{5.48})
and their reversed equations are consistent. However, for those $A$-values
for which $(\gam^A)^\ddg = - \gam^A$, the terms in the equations vanish.
The terms odd under reversion do not contribute to the equations of motion.
In this respect the situation corresponding to the action (\ref{5.48d})
is the same as that of the action (\ref{5.40}). The difference is that
in the case of the action (\ref{5.40}) there are two classes of
equations of
motion, while in the case of the action (\ref{5.48d}) the class with
odd $\gam^A$'s is missing.

This problem is avoided if we {\it redefine} the Clifford basis so
that insteaad of
\be
     \lbrace \gam_A \rbrace = \lbrace \gam_{a_1 ... a_r} \rbrace \; , 
     \quad r=1,2,...,n
\lbl{ 5.49a1}
\ee
we have
\be
     \lbrace \gam_A \rbrace = 
    \lbrace i^{\,r(r-1)/2}\, \gam_{a_1 ... a_r}  \rbrace
\lbl{5.49a2}
\ee
where $\gam_{a_1 ... a_r}$, $r=1,2,...,n$, are defined as the antisymmetrized
products (\ref{2.3}). Explicitly, for $n=4$, we have
\be
    \lbrace \gam_A \rbrace = \lbrace {\bf 1},\gam_{a_1},i \gam_{a_1 a_2},
    -i \gam_{a_1 a_2 a_3}, - \gam_{a_1 a_2 a_3 a_4} \rbrace
\lbl{5.49a3}
\ee
{\it The newly defined basis elements} are invariant under reversion:
\be
     \gam_A^\ddg = \gam_A
\lbl{5.49a4}
\ee
Using eq.\,(\ref{2.11}) we also have that the coordinate basis elements
are invariant under reversion:
\be
    \gam_M^\ddg = {e_M}^A \gam_A^\ddg = {e_M}^A \gam_A = \gam_M
\lbl{5.40a5}
\ee
Similarly for $\gam^A$ and $\gam^M$. Then we have that the square of the
generalized Dirac operator is equal to the generalized Klein-Gordon
operator:
\be
     \p \p \Psi = \gam^A \gam^B \p_a \p_B \Psi =
     {\gam^A}^\ddg \gam^B \p_a \p_B \Psi = G^{AB} \p_a \p_B \Psi
\lbl{5.40a6}
\ee

If we define the $C$-space Dirac equation by using the new basis
(\ref{5.49a2}), then the reversed equation (\ref{5.48c}) coincides with
eq.\,(\ref{5.48b}); the actions (\ref{5.48d}) and (\ref{5.40}) become 
equivalent.

In eq.\,(\ref{5.48}) the quantity $\delta \Psi = \Psi'(X) - \Psi (X)$ is
the variation of the field at fixed point $X= x^M \gam_M$. Following the
usual procedure we introduce the total variation
\be
    {\bar \delta} \Psi = \Psi'(X') - \Psi(X) = \delta \Psi + \p_M \Psi \,
    \delta x^M
\lbl{5.49}
\ee
which takes into account the variation of the point $X$ as well. Assuming
that the equations of motion are fulfilled, and inserting eq.\,(\ref{5.49})
into eq.\,(\ref{5.48}), we obtain
\be
    \delta I = i \int \dd^{2^n} X \, \p_M \left [ \sqrt{|G|} (\Psi^\ddg
    \gam^M {\bar \delta} \Psi - \Psi^\ddg \gam^M \p_N \Psi \, \delta x^N)
    \right ]
\lbl{5.50}
\ee
Here
\be
   G^M \equiv i(\Psi^\ddg
    \gam^M {\bar \delta} \Psi - \Psi^\ddg \gam^M \p_N \Psi \, 
    \delta x^N )
\lbl{5.50a}
\ee
is the generator for the corresponding transformation of our
generalized Dirac like action in $C$-space.

Let us now consider the following transformation
\bear
    &\Psi'(X') = R \Psi (X) S = {\rm e}^{\mbox{$1\oo 4$} \Sigma_{AB} 
    \alpha^{AB}} \Psi (X)
    {\rm e}^{\mbox{$1\oo 4$} \Sigma_{CD} \beta^{CD}}&
\lbl{5.51}
\\
     &x'^M = {e_A}^M x'^A = {e_A}^M {L^A}_B x^B&
\lbl{5.52}
\ear
This are the general ``rotations" (Lorentz transformations) in $C$-space.
The corresponding infinitesimal transformations (for infinitesimal parameters
$\alpha^{AB}$ and $\beta^{AB}$) are
\bear
   &{\bar \delta} \Psi = \Psi'(X') - \Psi (X) = 
   \mbox{$1\oo 4$}\Sigma_{AB}\, 
   \alpha^{AB} \Psi(X)
   + \Psi (X) \mbox{$1\oo 4$} \Sigma_{AB} \, \beta^{AB}&
\lbl{5.53}
\\
   &\delta x^M = {e_A}^M {\epsilon^A}_B x^B&
\lbl{5.54}
\ear
Here ${L^A}_B$ is a ``Lorentz" transformation in the tangent $C$-space
at a point $X$, and $\epsilon^{AB} = - \epsilon^{BA}$ are infinitesimal
parameters.

Inserting eqs.\,(\ref{5.53}),(\ref{5.54}) into eq.\,(\ref{5.50a}) we
obtain
\be
    G^M = i \Psi^\ddg \gam^M \mbox{$1\oo 4$}(\Sigma_{AB} \alpha^{AB} 
    \Psi +
    \Psi \Sigma_{AB} \beta^{AB}) - 
    i\Psi^\ddg \gam^M (x_A \p_B - x_B \p_A) \Psi \epsilon^{AB}
\lbl{5.55}
\ee
where
\be
    x_A \p_B = {e_A}^N {e_B}^J x_N \p_J
\lbl{5.56}
\ee

Parameters $\epsilon^{AB}$ on the one hand, and $\alpha^{AB},~\beta^{AB}$
on the other hand, denote the same local transformation (``rotation") in
$C$-space. In order to obtain the relation between the two sets of
parameters, we proceed as follows.

According to eqs.\,(\ref{4.1}),(\ref{4.10}), a polyvector can be
written either in terms of the polyvector components $\phi^A$, or the
generalized spinor components $\psi^{\tl A}$. The two sets of components
are related as
\be
    \psi^{\tl A} = {H^{\tl A}}_A \phi^A \; , \qquad \phi^A = {H^A}_{\tl B}
    \psi^{\tl B}
\lbl{5.57}
\ee
where ${H^{\tl A}}_A = \langle {\xi^{\tl A}}^\ddg \gam_A \rangle_S$ and
${H^A}_{\tl A} = \langle {\gam^A}^\ddg \xi_{\tl A} \rangle_S$ satisfy
\be
   {H^{\tl A}}_A {H^A}_{\tl B} = {\delta^{\tl A}}_{\tl B} \; , \qquad
   {H^A}_{\tl A} {H^{\tl A}}_B = {\delta^A}_B
\lbl{5.57a}
\ee
The transformation of $\psi^{\tl A}$ then reads
\be
    \psi'^{\tl A} (X') = {H^{\tl A}}_A \phi'^A (X') = {H^{\tl A}}_A
    {L^A}_B \phi^B (X) = {H^{\tl A}}_A {L^A}_B {H^B}_{\tl B} 
    \psi^{\tl B} (X)
    = {U^{\tl A}}_{\tl B} \psi^{\tl B} (X)
\lbl{5.58}
\ee
where
\be {U^{\tl A}}_{\tl B} = {H^{\tl A}}_A {L^A}_B {H^B}_{\tl B}
\lbl{5.59}
\ee
The infinitesimal transformation is
\be
    \psi'^{\tl A} (X') = {H^{\tl A}}_A ({\delta^A}_B + {\epsilon^A}_B)
   {H^B}_{\tl B} \psi^{\tl B} (X)
\lbl{5.60}
\ee
or
\be
    {\bar \delta} \psi^{\tl A} = 
    {H^{\tl A}}_A {\epsilon^A}_B {H^B}_{\tl B} \psi^{\tl B}
\lbl{5.61}
\ee
Writing the indices ${\tl A},~{\tl B}$ in the form of double indices,
${\tl A}=\gam \delta,~{\tl B}= \alpha \beta$ (where the ``ideal"
indices $i,j$ are now replaced by $\alpha$, $\beta$, that is, they are
written by the same symbol as spinor indices), the latter relation can
be written as
\be
     {\bar \delta} \psi^{\gam \delta} = {H^{\gam \delta}}_A {\epsilon^A}_B
     {H^B}_{\alpha \beta} \psi^{\alpha \beta}
\lbl{5.61a}
\ee
On the other hand, the same infinitesimal transformation is given in 
eq.\,(\ref{5.53}), which in component form reads
\be
    {\bar \delta} \psi^{\gam \delta} = \mbox{$1\oo 4$}
    {(\Sigma_{AB})^\gam}_\alpha \alpha^{AB}
    \psi^{\alpha \delta} + \psi^{\gam \beta} 
    \mbox{$1\oo 4$}{(\Sigma_{AB})_\beta}^\delta\beta^{AB}
    = {T^{\gam \delta}}_{\alpha \beta} \psi^{\alpha \beta}
\lbl{5.62}
\ee
where
\be
   {T^{\gam \delta}}_{\alpha \beta} =
   \mbox{$1\oo 4$}{(\Sigma_{AB})^\gam}_\alpha \alpha^{AB} 
   {\delta^\delta}_\beta
   + {\delta^\gam}_{\alpha} \mbox{$1\oo 4$}{(\Sigma_{AB})_\beta}^\delta
   \beta^{AB}
\lbl{5.62a}
\ee
In matrix notation the above relations read
\be
    {\bar \delta} \psi = {\bf T} \psi \; \qquad 
    {\bf T} = \mbox{$1\oo 4$}{\bf \Sigma}_{AB} \otimes {\bf 1} \, \alpha^{AB}
    + {\bf 1} \otimes \mbox{$1\oo 4$}{\bf \Sigma}_{AB} \, \beta^{AB}
\lbl{5.62b}
\ee
Eq.\,(\ref{5.62}) can be directly compared with eq.\,(\ref{5.61a}) and
so we obtain
\be
    {H^{\gam \delta}}_A {\epsilon^A}_B {H^B}_{\alpha \beta} =
    \mbox{$1\oo 4$}{(\Sigma_{AB})^\gam}_\alpha \alpha^{AB} 
    {\delta^\delta}_\beta
   + {\delta^\gam}_{\alpha} \mbox{$1\oo 4$}{(\Sigma_{AB})_\beta}^\delta
   \beta^{AB} = {T^{\gam \delta}}_{\alpha \beta}
\lbl{5.63}
\ee
Using eqs.\,(\ref{5.57a}) we can isolate the ${\epsilon^A}_B$ occurring in
eq.\,(\ref{5.63}):
\bear
  {\epsilon^C}_D &=& {H^C}_{\gam \delta} \left [\mbox{$1\oo 4$} 
 {(\Sigma_{AB})^\gam}_\alpha \alpha^{AB} {\delta^\delta}_\beta
   + {\delta^\gam}_{\alpha} \mbox{$1\oo 4$}{(\Sigma_{AB})_\beta}^\delta
   \beta^{AB} \right ]{H_D}^{\alpha \beta} \nonumber \\
    &=& {H^C}_{\gam \delta}
   {T^{\gam \delta}}_{\alpha \beta} {H_D}^{\alpha \beta}
\lbl{5.64}
\ear

Alternatively, we can derive the above relation from
\be
   X' = x^A \gam'_A = x'^A \gam_A = x^A \, R \gam S = {L^A}_B x^B \gam_A
\lbl{5.65}
\ee
Multiplying the latter equation by ${\gam^C}^\ddg$ and taking the scalar
part, we find
\be
    x^A \langle {\gam^C}^\ddg \, R \gam_A S \rangle_S = {L^C}_B x^B
\lbl{5.66}
\ee
Since $x^A$ is arbitrary, we have
\be
    \langle {\gam^C}^\ddg \, R \gam_D S \rangle_S = {L^C}_D
\lbl{5.67}
\ee
For an infinitesimal transformation we obtain
\be
   \langle {\gam^C}^\ddg \mbox{$1\oo 4$}(\Sigma_{AB} \alpha^{AB} \gam_D
    + \gam_D \Sigma_{AB} \beta^{AB} ) \rangle_S = {\epsilon^C}_D
\lbl{5.68}
\ee
Using 
\be
    {\gam^C}^\ddg = {H^C}_{\tl C} {\xi^{\tl C}}^\ddg \; , \qquad
    \gam_D = {H_D}^{\tl D} \xi_{\tl D}
\lbl{5.69}
\ee
eq.\,(\ref{5.68} becomes
\be
    {\epsilon^C}_D = {H^C}_{\tl C}\mbox{$1\oo 4$} 
    \left ( \langle {\xi^{\tl C}}^\ddg
    \Sigma_{AB} \xi_{\tl D} \rangle_S \alpha^{AB} + 
    \langle {\xi^{\tl C}}^\ddg \xi_{\tl D} \Sigma_{AB} \rangle_S 
    \beta^{AB} \right ) {H_D}^{\tl D}
\lbl{5.70}
\ee
Using the ciclic property of the operation $\langle ~~ \rangle_S$, namely,
$\langle {\xi^{\tl C}}^\ddg \mbox{$1\oo 4$}\Sigma_{AB} \xi_{\tl D} 
\rangle_S  =
\langle \xi_{\tl D} \mbox{$1\oo 4$}\Sigma_{AB} {\xi^{\tl C}}^\ddg 
\rangle_S$, and writing 
\be
    {\tl C}= \gam \delta,~{\tl D} = \alpha \beta
\lbl{5.70a}
\ee
we obtain
\bear
    &\langle {\xi^{\tl C}}^\ddg \Sigma_{AB} \xi_{\tl D} \rangle_S =
    \langle {\xi^{\gam \delta}}^\ddg \Sigma_{AB} \xi_{\alpha \beta} \rangle_S
    = {{(\Sigma_{AB})}^\gam}_\alpha \, {\delta^\delta}_\beta&
\lbl{5.71} \\
    &\langle \xi_{\tl D} \Sigma_{AB} {\xi^{\tl C}}^\ddg \rangle_S =
    \langle \xi_{\alpha \beta} \Sigma_{AB} {\xi^{\gam \delta}}^\ddg \rangle_S
    = {\delta_\alpha}^\gam {{(\Sigma_{AB})}_\beta}^\delta&
\lbl{5.72}
\ear
Inserting the latter relation into eq.\,(\ref{5.70}) we see that it is the
same as eq.\,(\ref{5.64}).

Let us now return to the generator (\ref{5.55}). Sandwiching eq.\,(\ref{5.50a})
between the spinors ${\xi^\sigma}^\ddg$ and $\xi^\delta 
\equiv \xi^{\delta 1}$,
and taking the scalar part, we obtain for the first term
\bear
   i \langle {\xi^\sigma}^\ddg \Psi^\ddg \gam^M {\bar \delta} \Psi \, 
   \xi^\delta \rangle_S &=& 
   i \langle {\xi^\sigma}^\ddg \Psi^\ddg \xi^\rho \xi_\rho^\ddg \gam^M 
   \xi^\gam \, {\xi^\gam}^\ddg {\bar \delta} \Psi \, \xi^\delta \rangle_S
   \nonumber \\
   &=&i \psi^{*\sigma \rho } (\gam^M )_{\rho \gam} {\bar \delta} 
   \psi^{\gam \delta} = 
   i {\psi^*}^{\sigma \rho} (\gam^M )_{\rho \gam}
   {T^{\gam \delta}}_{\alpha \beta}
   \psi^{\alpha \beta}
\lbl{5.73}
\ear
where $ {T^{\gam \delta}}_{\alpha \beta} $ is given in eq.\,(\ref{5.62a}).

Let us now use relation (\ref{5.63}). Then for the matrix elements of
the generator (\ref{5.55})  we obtain
\be
   \langle {\xi^\sigma}^\ddg G^M \xi^{\delta} \rangle_S =
   i \psi^{*\sigma \rho} (\gam^M )_{\rho \gam} 
   {H^{\gam \delta}}_A \, {\epsilon^A}_B {H^B}_{\alpha \beta} \,
   \psi^{\alpha \beta}
   + i \psi^{*\sigma \rho} (\gam^M )_{\rho \gam}
   (x_A \DD_B - x_B \DD_A) \psi^{\gam \delta} \epsilon^{AB}
\lbl{5.74}
\ee

In both parts that contribute to the generator we have now the same
parameters $\epsilon^{AB}$ of an infinitesimal rotation in $C$-space.
If we require that under a transformation generated by $G^M$ the action
remains invariant, i.e., $\delta I = 0$, then $G^M$ is conserved
generator: $(1/\sqrt{|G|}) \p_M (\sqrt{|G|} G^M) = \DD_M G^M =0$.
Considering {\it flat} $C$-space, we may choose a coordinate system in
which metric $G_{MN}$ and ${\rm det} \, G_{MN} \equiv G \neq 0$ are constant,
so that the conservation law is simply $\p_M G^M = 0$. Since
the transformation parameters $\epsilon^{AB}$ are arbitrary we also have
\be
     \p_M J_{AB}^M = 0
\lbl{5.75a}
\ee
where
\be
    \langle {\xi^\sigma}^\ddg J_{AB}^M \xi^\delta \rangle_S =
    i \psi^{*\sigma \rho} (\gam^M )_{\rho \gam} \left [
    {H^{\gam \delta}}_A H_{B \alpha \beta} + (x_A \p_B - x_B \p_A)
    {\delta ^\gam}_\alpha {\delta^\delta}_\beta \right ] \psi^{\alpha \beta}
\lbl{5.75b}
\ee
is the {\it Noether current} belonging to our generalized Dirac 
action (\ref{5.40}). It consists of the generalized {\it spin part} and
the generalized {\it orbital part}, and is conserved in flat $C$-space.

Alternatively, we can express the $\epsilon^{AB}$ occurring in the
second term of the generator (\ref{5.57a}) in terms of $\alpha^{AB}$ and
$\beta^{AB}$ according to eq.\,(\ref{5.64}). The generator (\ref{5.55}) then
reads
\bear
    &&G^M = \; i \Psi^\ddg \gam^M \mbox{$1\oo 4$}(\Sigma_{AB} \alpha^{AB}
    \Psi  +
    \Psi \, \Sigma_{AB} \beta^{AB} ) \hs{9cm} \nonumber \\
    &&\hs{12mm} + \; i \Psi^\ddg \gam^M
    (x_C \p_D - x_D \p_C) \Psi 
    {H^C}_{\gam \delta} \, \mbox{$1\oo 4$} \left [ 
{(\Sigma_{AB})^\gam}_\alpha \alpha^{AB} {\delta^\delta}_\beta
   + {\delta^\gam}_{\alpha} {(\Sigma_{AB})_\beta}^\delta \,
   \beta^{AB} \right ] {H_D}^{\alpha \beta}
\lbl{5.75}
\ear
From the fact that parameters $\alpha^{AB}$ and $\beta^{AB}$ are arbitrary,
we read the corresponding Noether current. It consists of the left part
due to the {\it left transformations} (\ref{5.7})
\be
    J_{AB}^M ({\rm left}) = i \Psi^\ddg \gam^M \mbox{$1\oo 4$}\Sigma_{AB}\Psi
    + i \Psi^\ddg \gam^M (x_C \p_D - x_D \p_C) \Psi
    {H^C}_{\gam \beta} {\mbox{$1\oo 4$}(\Sigma_{AB})^\gam}_\alpha 
    H^{\alpha \beta D}
\lbl{5.76}
\ee
and the part dues to the {\it right transformations} (\ref{5.9})
\be
   J_{AB}^M ({\rm right}) = i \Psi^\ddg \gam^M \Psi 
   \mbox{$1\oo 4$}\Sigma_{AB} +
   i \Psi^\ddg \gam^M (x_C \p_D - x_D \p_C) \Psi
   {H^C}_{\alpha \delta} 
   {{\mbox{$1\oo 4$}(\Sigma_{AB})}_\beta}^\delta H^{\alpha \beta D}
\lbl{5.77}
\ee

In this approach the charges that generate gauge transformations are on the
same footing as the spin and orbital angular momentum. This provides a
natural unification of spin and charges, a program that has been
pursued for many years by Manko\v c \ci{Mankoc}, using different methods
and by employing extra spacetime dimensions.

\subsection{The gauge field potentials and gauge field strengths}

\subsubsection{Gauge field potentials as parts of $C$-space spin
connection}

Using eq.\,(\ref{5.1}) we can express the spin connection in terms of
the generators $\Sigma_{AB}$:
\be \Gam_M = {1\oo 4} \,{\Omega^{AB}}_N\, \Sigma_{AB} = {A_M}^A \gam_A \;
, \qquad {A_M}^A = {1\oo 4} \,{\Omega^{CD}}_N \, {f_{CD}}^A
\lbl{5.78}
\ee
The matrices representing the Clifford numbers $\Gam_M$ can be calculated
according to
\be
   \langle {\xi^{\tl A}}^\ddg \Gam_M \xi_{\tl B} \rangle_S  = 
   {{\Gam_M}^{\tl A}}_{\tl B} \equiv {\bf \Gam}_M
\lbl{5.79}
\ee
This is the spin connection entering the $C$-space Dirac equation written
in the forms (\ref{4.39})--(\ref{4.41}).

The $C$-space Dirac equation can be split according to
\be
   [ \bgam^\mu (\p_\mu + {\bf \Gam}_\mu ) + \bgam^{\bar M} (\p_{\bar M} +
   {\bf \Gam}_{\bar M})] \psi = 0
\lbl{5.80}
\ee
where $M=(\mu,{\bar M})$, and ${\bar M}$ assumes all the values except $M=
\mu = 0,1,2,3$. 

From eq.\,({\ref{5.78}) we read that the gauge fields $\Gam_M$ contain:

\ \ (i) {\it The spin connection} of the 4-dimensional gravity $\Gam_\mu^{(4)}=
{1\oo 8} \,{\Omega^{ab}}_\mu [\gam_a,\gam_b]$ 

\ (ii) {\it The Yang-Mills fields} ${A_\mu}^{\bar A} \gam_{\bar A}$, where
we have split the local index according to $A=(a,{\bar
A})$. For ${\bar A} = {\underline {\bf o}}$ (i.e., for the scalar) the latter
gauge field is just that of the U(1)
group.

(iii) The antisymmetric potentials 
${A_M^{\underline {\bf o}}} \equiv A_M = A_\mu,~A_{\mu \nu},~A_{\mu \nu \rho},
 ~A_{\mu \nu \rho \sigma}$, if we take indices $A={\underline {\bf o}}$
 (scalar) and $M= \mu,~\mu \nu,~ \mu \nu \rho,~\mu \nu \rho \sigma$.
 
(iv) Non abelian generalization of the antisymmetric potentials
$A_{\mu \nu ...}^a$ and $A_{\mu \nu ...}^{\bar A}$\,.

We see that the $C$-space spin connection contains all physically interesting
fields, including the antisymmetric gauge field potentials which
occur in string and brane theories.

A caution is in order here. From eq.\,(\ref{5.78}) it appears as if there
were only 16 independent generators $\gam_A$ in terms of which a gauge field
is expressed. But inspecting the generators 
(\ref{5.1}) we see that there are
more than 16 different rotations in $C$-space. However, some of them,
although being physically different transformations, turn out to be
mathematically described by the same objects. For instance, the generators
$\Sigma_{ab} = {1\oo 2} [\gam_a,\gam_b]$ and $\Sigma_{{\tl a}{\tl b}}
={1\oo 2} [\gam_5 \gam_a, \gam_5 \gam_b] = {1\oo 2} [\gam_a,\gam_b]$
are equal, although the corresponding transformations, i.e., a rotation
in the subspace $M_4$ and the rotation in the dual space ${\tl M}_4$ are
in principle independent. Similarly we have $\Sigma_{1 {\tl 1}} =
{1\oo 2} [\gam_1, \gam_5 \gam_1] = \gam_5 = \Sigma_{{\bf o}{\tl {\bf o}}}
= {\bf 1} \gam_5$. Such degeneracy of the transformations
is removed by the fact that the transformation
can act on the generalized spinor polyvector $\Psi$  either from the
left or from the right (according to (\ref{5.2})). 

Returning to the Dirac equation (\ref{5.80}) we see that besides the
part having essentially the same form as the ordinary Dirac equation in the
presence of minimally coupled 4-dimensional spin connection and 
Yang-Mills fields, there is also an extra term which can have the
role of mass,
if $\psi$ is an eigenstate of the operator $\gam^{\bar M} (\p_{\bar M}
+ \Gam_{\bar M})$. Since the metric signature of $C$-space is \ci{PavsicParis}
$(8+,8-)$, and the signature of the ``internal" space is $(7+,5-)$, the mass
is not necessarily of the order of the  Planck mass; it can be small due
to cancellations of the positive and negative contributions.

\subsubsection{Gauge field strengths as parts of $C$-space curvature}

Using eq.\,(\ref{4.38}) we can calculate the curvatue according to

\be
     [\p_M,\p_N] \xi_{\tl A}= {{R_{MN}}^{\tl B}}_{\tl A}\, \xi_{\tl B}
\lbl{5.81}
\ee
where
\be
    {{{R_{MN}}^{\tl B}}}_{\tl A} = \p_M {{\Gam_N}^{\tl B}}_{\tl A} -
    \p_N {{\Gam_M}^{\tl B}}_{\tl A} + 
    {{\Gam_M}^{\tl B}}_{\tl C} {{\Gam_N}^{\tl C}}_{\tl A} -
    {{\Gam_N}^{\tl B}}_{\tl C} {{\Gam_M}^{\tl C}}_{\tl A}
\lbl{5.81a}
\ee
This is the relation for the Yang-Mills field strength. In matrix
notation it reads:
\be
   {\bf R}_{MN}= \p_M \boldsymbol{\Gamma}_N - \p_N \boldsymbol{\Gamma}_M
   + [\boldsymbol{\Gam}_M,\boldsymbol{\Gam}_N]
\lbl{5.82a}
\ee
Inserting eqs.(\ref{5.78}),(\ref{5.79}) into eq.(\ref{5.82a}) we obtain
\be
   {F_{MN}}^A = \p_M {A_N}^A - \p_N {A_M}^A + {A_M}^B {A_N}^C {C_{BC}}^A
\lbl{5.82b}
\ee
where ${C_{BC}}^A$ are the structure constants of the Clifford algebra:
\be
        [\gam_A,\gam_B] = {C_{AB}}^C \gam_C
\lbl{5.82c}
\ee

From the curvature
we can form the invariant expressions, for instance
\be {R_{MN}}^{{\tl A} {\tl B}}
({\gam^M}^\ddg * \xi_{\tl A})({\gam^N}^\ddg * \xi_{\tl B})
={R_{MN}}^{{ A} { B}}
({\gam^M}^\ddg * \gam_{ A})({\gam^N}^\ddg * \gam_{ B})
= {R_{MN}}^{{ A} { B}}\, {e^M}_A \, {e^N}_B = R
\lbl{5.83}
\ee 
which is linear in curvature, and
\be
    {R_{MN}}^{{\tl A} {\tl B}}{R^{MN}}_{{\tl A} {\tl B}} =
     {R_{MN}}^{AB} \, {R^{MN}}_{AB}
\lbl{5.84}
\ee
which is quadratic in curvature. Instead of the form (\ref{5.84})
we can use (\ref{5.82b}) and take its square:
\be
      {F_{MN}}^A {F^{MN}}_A
\lbl{5.85}
\ee
{\it The action} for our system thus contains the term (\ref{5.40}) for the
generalized Dirac field $\psi^{\tl A}$ and the kinetic term for gauge fields:
\be
    I[{A_M}^A] = \int \dd x^{2^n} \,\sqrt{|G|}\, (\alpha R + 
    \beta {F_{MN}}^A {F^{MN}}_A)
\lbl{5.86}
\ee
Here $\alpha$ and $\beta$ are the coupling constants. The first term
in the action (\ref{5.86}) is a generalization of the Einstein gravity,
whereas the second term generalizes the higher derivative gravity to
$C$-space. Other terms of the $R^2$ type can also be added to the action
(\ref{5.86}).

\subsubsection{Conserved charges and isometries}

In curved space in general there are no conserved quantities, unless
there exist isometries which are described by Killing vector fields.
Suppose we have a curved Clifford space which admits $K$ Clifford numbers
$k^\alpha = k_M^\alpha\, \gam^M\,$,  ~$\alpha = 1,2,...,K;~ M=1,2,...,16$,
where the components satisfy the condition for isometry, namely
\be
     \DD_N k_M^\alpha + \DD_M k_N^\alpha = 0
\lbl{5.87}
\ee
the covariant derivative being defined in eq.\,(\ref{5.15}). 
We assume that such $C$-space with isometries is not given ad hoc, but
is a solution to the generalized Einstein equations that arise from the action
which contains the ``matter" term (\ref{5.40}) and the field term (\ref{5.86}).

Taking a coordinate system in which $k^{\alpha \mu}=0,~k^{\alpha {\bar M}}
\neq 0,~\mu=0,1,2,3,~{\bar M} \neq \mu$, the metric and
vielbein can be written as\footnote{This is a $C$-space analog of the
Kaluza-Klein splitting usually performed in the literature. See, e.g.,
\ci{Luciani,Witten}.}
\be
   G_{MN} = \begin{pmatrix}
            G_{\mu \nu} & G_{\mu {\bar M}}\\
            G_{{\bar M} \nu} & G_{{\bar M}{\bar N}}\\
            \end{pmatrix} \; , \quad 
    {e^A}_M = \begin{pmatrix}
            {e^a}_\mu & {e^a}_{\bar M} \\
            {e^{\bar A}}_\mu & {e^{\bar A}}_{\bar N} \\
            \end{pmatrix} 
\lbl{5.88}
\ee
Here
\be
    {e^a}_{\bar M} = 0
\lbl{5.89}
\ee
whilst the components ${e^{\bar A}}_\mu$ can be written in terms of Killing
vectors and gauge fields $W_\mu^\alpha (x^\mu)$:
\be
    {e^{\bar A}}_\mu = {e^{\bar A}}_{\bar M} \, 
    k^{\alpha {\bar M}} W_\mu^\alpha \; , \quad
    \quad \p_{\bar M} W_\mu^\alpha = 0
\lbl{5.90}
\ee
If we set the $C$-space torsion to zero and calculate
the connection $\Omega_{ABM}$, given in eq. (\ref{2.29}), by
using eqs.\,(\ref{5.88})--(\ref{5.90}), we obtain an analogous result
as given, e.g., in ref.\,\ci{Luciani}:
\be
    \Omega_{{\bar M}{\bar N} \, \mu} = 
    {1\oo 2} k_{[{\bar M},{\bar N}]}^\alpha W_\mu^\alpha
\lbl{5.91}
\ee
where $ k_{[{\bar M},{\bar N}]}^\alpha \equiv
\p_{\bar N} k_{\bar M}^\alpha -\p_{\bar M} k_{\bar N}^\alpha$

Let us now rewrite the $C$-space Dirac equation by using
eqs.\,(\ref{5.87})--(\ref{5.91}). Omitting the terms due to the
$C$-space torsion, we obtain
\be
    \left [ {{\bf \gam}^{(4)}}^\mu 
    \left ( \p_\mu - \Omega_{ab \,\mu} {1\oo 8} 
    [{\bf \gam}^a,{\bf \gam}^b] - q^\alpha \, W_\mu^\alpha
     + ...\right ) + 
    {\bf \gam}^{\bar M} \p_{\bar M} + ... \right ] \psi
= 0
\lbl{5.92}
\ee
where ${{\bf \gam}^{(4)}}^\mu = {e_a}^\mu {\bf \gam}^a$ are 4-dimensional
basis vectors in coordinate frame, and
\be
     q^\alpha = k^{\alpha \, {\bar M}} \, \p_{\bar M} + {1\oo 8}\, 
      k_{[{\bar M},{\bar N}]}^\alpha \,{e_{\bar A}}^{\bar M}
      {e_{\bar B}}^{\bar N}\, \Sigma^{{\bar A}{\bar B}}
\lbl{5.93}
\ee
are the charges, conserved due to the presence of isometries
$k^{\alpha \, {\bar M}}$. They are the sum of the coordinate
part and the contribution of the spin angular momentum in the ``internal"
space, spanned by $\gam^{\bar M}$. The coordinate part is the projection
of the linear momentum onto the Killing vectors, and can in particular be
just the orbital angular momentum of the ``internal" part of $C$-space.
The first term that contributes to the charge $q^\alpha$ comes from
the vielbein  according to eq.\,(\ref{5.90}), whilst the second term comes
from the connection according to eq.\,(\ref{5.91}). Both terms couple
to the same
4-dimensional gauge fields $W_\mu^\alpha$, where the index
$\alpha$ denotes which gauge field (which Killing vector), and should not
be confused with the spinorial index, used in Dirac matrices.

In eq.\,(\ref{5.92}) we explicitly displayed only the the most relevant
terms which contain the the ordinary
vierbein ${e^a}_\mu$ and spin connection $\Omega_{ab \,\mu}$
(describing gravity and torsion), and also Yang-Mills gauge fields
$W_\mu^\alpha$ which, as shown in eqs.\,(\ref{5.90}),(\ref{5.91}), occur
in the $C$-space vielbein and in the $C$-space ``spin" connection.
We omitted the terms arising form the $C$-space torsion

\section{Discussion}

A motivation for this study is a very promising possibility that the
generalized spinors (polyvectors)
$\Psi = \psi^{\tl A} \xi_{\tl A}$ provide a
representation of the gauge group U(1)$\times$SU(2)$\times$SU(3) of the
standard model, and that this group is incorporated in the group
GL(4,C)$\times$GL(4,C) formed by the transformations (\ref{5.2}).
The general spinors $\Psi$ in fact form a representation of the group
GL(4,C)$\times$GL(4,C) which is further restricted by the requirement
that the quadratic form $\langle \Psi^\ddg \Psi \rangle_S = 
{\psi^*}^{\tl A} Z_{{\tl A}{\tl B}} \psi^{\tl B}$ should be invariant under
the transformations (\ref{5.2}). The generalized spinor metric 
$Z_{{\tl A}{\tl B}} = \langle \xi_{\tl A}^\ddg \xi_{\tl B} \rangle_S$
encompasses sixteen basis spinors, four for each left ideal.
If the basis spinors are constructed according to the lines as
suggested in
eqs.\,(\ref{4.2})--(\ref{4.9}), one finds that there are several possibilities,
giving different signatures of the spinor metric. Two cases of possible
signature that arise in such a construction are of particular interest:

Case \; ~(i):
\be
   {\rm Sig} (Z_{{\tl A} {\tl B}}) = 
    \begin{pmatrix} + & + & - & - & \\
            + & + & - & - & \\
            - & - & + & + & \\
            - & - & + & + & 
     \end{pmatrix}
\lbl{4B.1}
\ee

Case \; (ii):
\be
   {\rm Sig} (Z_{{\tl A} {\tl B}}) = 
    \begin{pmatrix} + & - & - & - & \cr
            + & - & - & - & \cr
            - & + & + & + & \cr
            - & + & + & + & \cr \end{pmatrix}
\lbl{4B.2}
\ee
We see that, while in Case (i) the right ideals have signature $(++--)$
or $(--++)$, whilst in Case (ii) the right ideals have signature $(+---)$
or $(-+++)$.

In {\it Case (i)} the group of the left transformations ${\bf R}\in
\lbrace{\bf R} \rbrace $ contains
U(2,2), and the group of the right transformations
${\bf S}\in\lbrace{\bf S} \rbrace $ contains
U(2,2) as well.

In {\it Case (ii)} the group of the left transformations still contains U(2,2),
whereas the group of the {\it right transformations} contains U(1,3).

A subgroup SU(3) naturally occurs in U(1,3), just as SO(3) occurs
within the Lorentz group SO(1,3). We have here a possibility of associating
leptons with the first left ideal, and three color states of quarks
with the remaining three left ideals (columns). The group SU(3), which operates
from the right, mixes three color states. In addition there occurs
a transformation,
analogous to a {\it boost}, and it mixes leptons and quarks.
However, such  transitions, according to our model, cannot occur
spontaneously; we expect that they require a huge amount of energy,
and also higher grade components of momentum,
therefore normally they cannot be observed\footnote{
In ref.\,(\ci{PavsicArena}) we demonstrated that  boosts mixing,
for instance, vector and bivector or threevector coordinates indeed
require large amounts of energy.}.

Our total group
$\lbrace{\bf R} \rbrace \times \lbrace{\bf S} \rbrace$, in {\it Case (ii)},
contains subgroups
SL(2,C) and U(1) $\times$ SU(2) $\times$ SU(3).
The former group describes the Lorentz
transformations in Minkowski space (which is a subspace of
flat $C$-space\footnote{If $C$-space manifold is curved, then we consider
one of its tangents spaces $T_X (C)$, everyone of which contains
Minkowski space as a subspace.}),
whilst the latter group coincides with the gauge group of the standard
model which describes electroweak and strong interaction.  
Whether this indeed provides a description of the standard model
remains to
be fully investigated. But there is further evidence in favor of the above
hypothesis in the fact that a polyvector field $\Psi = 
\psi^{\tl A} \xi_{\tl A}$
has 16 complex components. Altogether it has 32 real components. This
number matches, for one generation, the number of independent states for
spin, weak isospin and color (together with the corresponding antiparticle
states) in the standard model. A complex polyvector
field $\Psi$ has thus enough degrees of freedom to form a representation
of the group GL(4,C)$\times$ GL(4,C) which contains the Lorentz group
SL(2,C) and the group of the standard model 
U(1)$\times$SU(2)$\times$SU(3). The generators of the group are given
by $\Sigma_{AB}$ defined in eq. (\ref{4.32}).

In our model, assigning the generalized spin metric signature (\ref{4B.2}),
we do not need artificially kill pieces of an SU(2n) in order to
obtain SU(3). Instead, we have SU(1,3) operating from the right, and SU(3)
is a natural subgroup.
Since starting from a different model, to obtain SU(3) most authors have
to do very unmotivated ``turning-off" of degrees of freedom. Chisholm
and Farwell \ci{Chisholm} do have polyvector wavefunctions, but they do
not fully utilize all the possible degrees of freedom. Rather they
simulate a column spinor, with the rest of the matrix filled with zero.
Instead they consider extra dimensions of spacetime. Hence they ended
up with a really big Clifford algebra operating only from the left.
A different approach, using octonions, is proposed by Dixon \ci{Dixon},
and recently by Dray and Manogue \ci{Dray}.

\section{Conclusion}

Spacetime manifold $V_4$ can be elegantly described by means of the basis
vectors which are generators of Clifford algebra. The latter algebra
describes a geometry which goes beyond spacetime: the ingredients
are not only points, but also 2-surfaces, 3-volumes, 4-volumes
and scalars. All those geometric objects form a 16-dimensional
manifold, called Clifford space, or shortly, $C$-space. It is quite
possible that the arena for physics is not spacetime, but Clifford
space. And the arena itself can become a part of the play,
if we assume that Clifford space is curved, and that its curvature
is a dynamical quantity entering the action functional. We have
thus a higher dimensional curved differential manifold.
From now on we can proceed \` a la Kaluza-Klein. Since the ``extra
dimensions'' are assumed to be related to the physical degrees of
freedom, due to the extended nature of physical objects, there is no
need to compactify the 12-dimensional ``internal'' part of
$C$-space. 

The theory that we pursue here has not only the prospects for providing
a clue to the unification of fundamental interactions. It provides
a framework for a generalized relativistic dynamics, including generalized
gravity, which might find its useful applications in astrophysics
and cosmology, which are fast developing fields, where many
surprises has already taken place, and more are to be expected on
the way.

\vs{5mm}  

\centerline{Acknowledgement}

This work was supported by the Ministry of
High Education, Science and Technology of Slovenia

{\small

\end{document}